\documentclass[pra,aps,amsmath,amssymb,amsfonts,twocolumn,nofootinbib,longbibliography,superscriptaddress]{revtex4-1}
\usepackage{amssymb}
\usepackage{bm,mathrsfs}
\usepackage{graphicx}
\usepackage{epsfig}
\usepackage{amsmath,bbm}
\usepackage{amsfonts,amssymb}
\usepackage{times}
\usepackage{verbatim}
\usepackage[sort&compress]{natbib}
\usepackage{amsmath}
\usepackage{bm}
\usepackage{float}
\usepackage{mathtools}
\usepackage{ragged2e}
\allowdisplaybreaks[4]
\usepackage[colorlinks,breaklinks,linkcolor=blue,anchorcolor=blue,citecolor=blue,urlcolor=magenta]{hyperref}

\usepackage{color}
\definecolor{MyDarkBlue}{rgb}{0,0.08,0.45}
\definecolor{yellow}{rgb}{0.99,0.99,0.70}
\definecolor{white}{rgb}{1.0,1.0,1.0}
\definecolor{black}{rgb}{0.00,0.00,0.00}
\definecolor{green}{rgb}{0.8,0.98,0.83}

\begin{document}
\title{Higher-Order Exceptional Points Induced by Non-Markovian Environments}

\author{H. Z. Shen}
\email{Corresponding author: shenhz458@nenu.edu.cn}
\affiliation{Center for Quantum Sciences and School of Physics,
Northeast Normal University, Changchun 130024, China}

\author{X. C. Zhang}
\affiliation{Center for Quantum Sciences and School of Physics,
Northeast Normal University, Changchun 130024, China}

\author{L. Y. Ning}
\affiliation{Center for Quantum Sciences and School of Physics,
Northeast Normal University, Changchun 130024, China}

\author{Zhi-Guang Lu}
\affiliation{School of Physics, Huazhong University of Science and Technology,
Wuhan 430074, People’s Republic of China}

\author{Yan-Hui Zhou}
\email{Corresponding author: yanhuizhou@126.com}
\affiliation{Quantum Information Research Center and Jiangxi Province Key Laboratory of Applied Optical Technology,
Shangrao Normal University, Shangrao 334001, China}

\author{Cheng Shang}
\email{Corresponding author: cheng.shang@riken.jp}
\affiliation{Analytical Quantum Complexity RIKEN Hakubi Research Team,
RIKEN Center for Quantum Computing (RQC), Wako, Saitama 351-0198, Japan}
\affiliation{Department of Physics, The University of Tokyo,
5-1-5 Kashiwanoha, Kashiwa, Chiba 277-8574, Japan}

\begin{abstract}
    Exceptional points (EPs) are central to non-Hermitian physics because of their unique properties and broad application prospects. While extensively studied in parity-time ($\mathcal{P}\mathcal{T}$)-symmetric systems and under Markovian dynamics, their exploration in broader pseudo-Hermitian settings, particularly in those involving non-Markovian environments, remains largely unexplored. In this study, we investigate a pseudo-Hermitian system consisting of three coupled optical cavities interacting with non-Markovian environments. Compared to the Markovian baseline, we demonstrate that the emergence of non-Markovian memory effects enlarges the dimensionality of the parameter space of the system, thereby giving rise to higher-order EPs. Furthermore, we show that these non-Markovianity-induced higher-order EPs admit a topological characterization as defects in the relevant pseudo-Hermitian parameter space, with quantized charges described by resultant winding numbers. Moreover, we observe that the pseudo-Hermitian system with an effective gain induced by coherent perfect absorption enables the higher-order EPs to be directly read out from the output spectrum. We also note that the non-Markovian mechanism for generating higher-order EPs is not restricted to pseudo-Hermitian systems, but can be extended to generic non-Hermitian quantum systems. Additionally, possible experimental implementations based on superconducting circuits are also discussed.
\end{abstract}

\maketitle


\section{Introduction}
Superconducting (SC) circuits have emerged as a promising platform for quantum information processing and for investigating unique physical phenomena, as extensively reviewed in Ref.~\cite{Xiang6232013}. Quijandr\'{\i}a \emph{et al}.~introduced the concept of a $\mathcal {P}\mathcal {T}$-symmetric phase transition occurring at the EP in SC circuits~\cite{8462018}, which was subsequently observed on SC quantum computing platform of IBM by Dogra \emph{et al}.~\cite{Dogra262021}. Following these theoretical advances, a series of experiments have verified EP-related phenomena in SC platforms, with EP signatures detected in dissipative SC qubits~\cite{Naghiloo2322019,Chen5042021,Chen4022022} and in interconnected systems comprising two dissipative SC resonators~\cite{Partanen5052019}. Moreover, Han \emph{et al}. have experimentally demonstrated the exceptional entanglement transition in the vicinity of an EP~\cite{Han2012023}, alongside the topological invariant associated with EP3~\cite{Han02839}, through meticulous monitoring of the dynamical behavior within SC circuits. Building on this progress, recent work by Zhang \emph{et al}. has further advanced the field by studying the higher-order exceptional surface in SC circuits~\cite{Zhang06062}.

Over the past decades, EPs have attracted remarkable attention~\cite{Feng7522017,Ganainy112018,Ozdemir7832019,Wiersig0052021,Bergholtz0052021,Zhang020328,Khandelwal04036,Huang2100430}. In particular, a $k$th-order EP (EP$k$, with $k \geq 2$) refers to a spectral singularity of a non-Hermitian Hamiltonian where $k$ eigenvalues and their corresponding eigenstates coalesce~\cite{Heiss0162012,Schnabel053868,Chakraborty063846,Laha063829,Ramezanpour043510,Mandal186601}. In non-Hermitian systems, the spectral anomaly associated with EPs has been demonstrated to induce a multitude of fascinating phenomena, including unidirectional invisibility~\cite{Lin9012011,Peng3942014}, resilient wireless energy transmission~\cite{3872017,Hao2022023}, asymmetric modal transitions~\cite{Doppler762016,Xu802016}, augmented spontaneous radiation~\cite{Lin4022016}, unidirectional lasing~\cite{Longhi2017}, exotic topological states~\cite{Zhou0092018}, sensitivity enhancement~\cite{Hodaei1872017,Wiersig9012014,Chen1922017,Zhang3072021,Liu8022016}, laser mode selection~\cite{Feng9722014,Hodaei9752014}, coherent perfect absorption (CPA)~\cite{Sun9032014,Zhang3682017,Wang2612021}, electromagnetically induced transparency~\cite{Wang3342020}, the speeding up of entanglement generation~\cite{Li2022023}, and the exhibition of complex topological characteristics in interconnected acoustic resonators~\cite{Ding0072016}.

Closed quantum systems satisfy Hermiticity and have purely real spectra.~By contrast, realistic systems are inevitably open and require an effective description in terms of non-Hermitian Hamiltonians with complex spectra.~Nevertheless, the reality of eigenvalues does not require Hermiticity; in other words, Hermiticity is not a necessary condition for a spectrum to be real. In this sense, Mostafazadeh identified a class of quantum systems exhibiting the above energy spectral properties, whose Hamiltonians are related to their adjoints through a similarity transformation $\eta $ given by $H = {\eta ^{ - 1}}{H^\dag }\eta $, known as pseudo-Hermitian Hamiltonians~\cite{Mostafazadeh2052002,Mostafazadeh22002,Mostafazadeh32002}, where eigenvalues are constrained to be either purely real or to appear in complex-conjugate pairs. Nowadays, systems possessing pseudo-Hermiticity have given rise to numerous interesting phenomena across diverse areas of physics, including quantum chaos, quantum phase transitions~\cite{Deguchi1072009,Deguchi2132009,Shen0121072016} and dynamical invariants \cite{Simeonov1232016}.

In particular, the Hermitian Hamiltonians form a subset of pseudo-Hermitian ones, while $\mathcal{P}\mathcal{T}$-symmetric Hamiltonians constitute another prominent subclass within the pseudo-Hermitian landscape~\cite{Mostafazadeh32002,Konotop0022016}. A quantum phase transition occurs as system parameters approach the EP, driving the system from a $\mathcal{P}\mathcal{T}$-symmetric to a symmetry-broken phase. This transition is characterized by the spectrum shifting from real to complex eigenvalues~\cite{Konotop0022016}, with the critical point termed an EP2, which has been extensively investigated in various non-Hermitian systems, including optomechanical setups \cite{L¨¹6012015,Xu802016}, coupled waveguides \cite{Doppler762016}, optical microresonator networks \cite{Chang5242014}, cavity magnonics systems \cite{Gao8262017}, and superconducting circuit-QED configurations \cite{8462018}. While $\mathcal{P}\mathcal{T}$ symmetry provides a fertile ground for exploring EPs, its stringent parameter constraints—especially for higher-order EPs—limit its scope. This motivates the broader framework of pseudo-Hermiticity, which encompasses both Hermitian and $\mathcal{P}\mathcal{T}$-symmetric cases and supports a wider range of non-Hermitian phenomena.~Within this framework, higher-order EPs and their applications have been explored on various platforms, including cavity-magnon systems \cite{Zhang4042019,Zhang4172023,Hu01613}, cavity optomechanical systems \cite{Xiong5082021,Xiong5182022}, and radio-frequency circuits \cite{Yin0032023}.

The rapid advances in quantum information technology \cite{nielsen2000} have brought open quantum systems \cite{BreuerOxford2002,White020344,Figueroa-Romero040351,GardinerBerlin2000,Li0621242010,Caruso12032014,breuer880210022016,Vega0150012017} into sharp focus. The Markovian approximation for open quantum systems \cite{BreuerOxford2002,Milz030201,Gribben010321,Mann030321,Diosi0341012012} applies only in the regime of weak system–environment coupling, where the characteristic timescales of the system significantly exceed the environmental correlation time. In contrast, many scenarios demand a full consideration of non-Markovian dynamics \cite{USP-NM2018}, which arise in diverse quantum setups such as interconnected cavities \cite{Link0203482022}, photonic crystals \cite{hoeppe1080436032012}, colored-noise environments \cite{CostaFilho0521262017}, cavity–waveguide hybrids \cite{Longhi0638262006,Sounas2017,Vega0150012017,Shen0437142024}, etc.~\cite{Yang053712,Liu20117,Xiong2019100,Cialdi2019100,Khurana201999,Madsen2011106,Guo2021126,Li2022129,Xu201082,Uriri2020101,Anderson199347,Liu2020102,Groblacher76062015}. Such non-Markovian processes have proven crucial for a variety of quantum information tasks, including state engineering and quantum control \cite{Shen0321012019,Xin0537062022}.~Their hallmark is the appearance of memory effects, wherein the influence of an environment on system dynamics reaches back to affect earlier states. Physically, this is often manifested as the repeated exchange of excitations between the system and its surroundings \cite{breuer880210022016,Shen7072022}, which in turn forms the foundation for a variety of approaches to quantifying non-Markovianity \cite{lorenzo880201022013,rivas1050504032010,luo860441012012,wolf1011504022008,lu820421032010,chruscinski1121204042014}.

Before proceeding, we place our results in the broader development of non-Markovian EPs. Early studies showed that memory effects can qualitatively reshape non-Hermitian spectral structures, for example through anomalous-order EPs and non-Markovian Purcell effects near structured continua \cite{Garmon033029}, as well as through the interplay between non-Markovian dissipation, quantum-Zeno-related dynamics, and EP behavior in open quantum systems \cite{Mouloudakis053709}. Time-delayed anti-\(\mathcal{PT}\)-symmetric systems further demonstrated that memory effects can enrich the EP landscape by effectively introducing an infinite-dimensional dynamical structure \cite{Wilkey1299}. More recently, general frameworks based on pseudomode mapping and hierarchical equations of motion have shown that structured environments can supply auxiliary degrees of freedom and generate EPs that are absent in the Markovian approximation \cite{Lin18362}. This viewpoint has been supported experimentally in superconducting-circuit Liouvillian systems \cite{HaoLZ2025}, and has been extended to revival-induced high-order dynamical EP signatures in finite-size reservoirs \cite{Sergeev251209582}, interpolated quantum channels \cite{Wong1263}, and waveguide-QED platforms with retardation and feedback effects \cite{Longhi70277}.


Our work builds on these developments but addresses a different and complementary question: how structured non-Markovian reservoirs can be used as controllable resources for engineering higher-order EPs in a pseudo-Hermitian optical system. We demonstrate that non-Markovianity induces higher-order EPs in a pseudo-Hermitian system comprised of three coupled optical cavities, and further clarify the underlying physical mechanisms involved. In our construction, distinguishable Lorentzian spectral components act as auxiliary pseudomode dimensions of the effective Hamiltonian. This reservoir-induced enlargement provides a direct route from the lower-order EP structure of the Markovian system to tunable higher-order EPs, ranging from EP4~\cite{CrippaL121109,Bhattacherjee062124,Jin012121,Othman104305,Dey07903,Arkhipov012205,Zhou10118,Shen065102,Kaltsas4447,Zhang2020} and EP5~\cite{Bai266901,Pan063616,Zhang033820} to EP6~\cite{Xiao213901,Kullig2023} and EP7, which are not present in Markovian models where the EP order is limited by the original system size. We further derive the corresponding pseudo-Hermiticity constraints and show that these higher-order EPs can be directly read out from the total output spectrum under CPA. We also discuss the extension of this non-Markovian mechanism for inducing higher-order EPs beyond the pseudo-Hermitian case to generic non-Hermitian quantum systems, and the experimental feasibility of observing higher-order EPs in superconducting circuits~\cite{Gu12017,Partanen5052019,Baust5152015,Zhang2022020,Almanakly2025,Nie2023,Kannan2023,HaoLZ2025}. Our work lays a solid foundation for exploring the spectral structures of higher-order EPs induced by non-Markovian environments and for investigating their potential applications near the EPs, such as highly sensitive sensors~\cite{Qin20005692021,Li0335052021,Jiang2011062023,Zhong0132202021,Soleymani5992022,Tang66602023,Jia2932023, Stalhammar2011042021,Chen0315012022,Zhong0140702020}.

The present paper is organized as follows. In Sec.~\ref{section II}, we introduce our model and its non-Markovian effective Hamiltonian. In Sec.~\ref{section III}, we demonstrate and classify non-Markovian higher-order EPs in both pseudo-Hermitian and generic non-Hermitian scenarios, and discuss the topological invariants associated with  higher-order EPs.~Possible experimental implementations are presented in Sec.~\ref{experimental analysis}. Finally, we summarize the results and conclude the paper in Sec.~\ref{Discussion and conclusions}.

\section{Model and Hamiltonian} \label{section II}

Since physical systems are inevitably open, non-Markovian effects are broadly relevant. While Markovian processes can capture many quantum phenomena, this approximation fails when the system strongly interacts with the environment and the environment memory time is comparable to the system characteristic time~\cite{GardinerBerlin2000,Li0621242010,Caruso12032014}, at which point non-Markovian effects dominate. We start this section by describing our model. Subsequently, we present the effective Hamiltonian under the condition for CPA.

\begin{figure}[h]
    \centering
    \includegraphics[width=0.3 \textwidth]{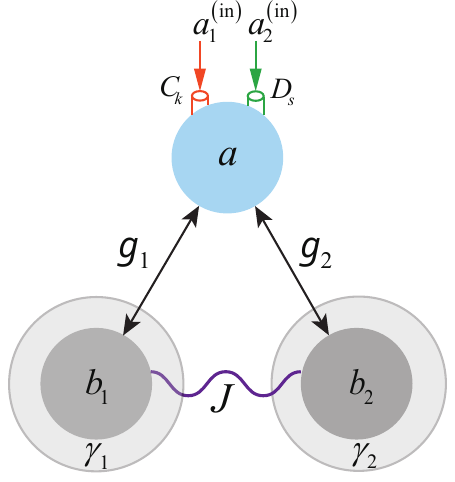} 
    \caption{Schematic of a three-mode optical cavity coupling setup in an open quantum system. Cavity mode $a$ couples to cavity modes $b_{1}$ and $b_{2}$ with strengths $g_{1}$ and $g_{2}$, respectively. Meanwhile, cavity modes $b_{1}$ and $b_{2}$ are coupled with strength $J$. In addition, the input fields $a_{1}^{(\mathrm{in})}$ and $a_{2}^{(\mathrm{in})}$ generate non-Markovian baths, which couple to cavity mode $a$ through two ports with strengths $C_{k}$ and $D_s$, respectively. Furthermore, cavity modes $b_{1}$ and $b_{2}$ are each coupled to Markovian environments with dissipation rates $\gamma_{1}$ and $\gamma_{2}$.}
    \label{fig:setup}
\end{figure}

\textit{Model}. We consider a three-mode optical open quantum system as illustrated in Fig.~\ref{fig:setup}. The central cavity mode $a$ couples to two side cavities $b_1$ and $b_2$ with interaction strengths $g_1$ and $g_2$, respectively, while the two side cavities are coupled to each other with strength $J$. Moreover, cavity $a$ is driven by two input fields $a_{1}^{(\mathrm{in})}$ and $a_{2}^{(\mathrm{in})}$ through two ports, which are equivalent to being coupled to two structured non-Markovian reservoirs, each consisting of a series of bosonic modes ${c_k}$ and ${d_s}$, with coupling strengths $C_k$ and $D_s $, respectively. On the other hand, the side cavities $b_1$ and $b_2$ interact with their respective Markovian reservoirs, characterized phenomenologically by dissipation rates $\gamma_1$ and $\gamma_2$. We assume that the two input fields have the same frequency $\omega_p$. In the rotating frame with $\omega_p$, the total Hamiltonian can be written as
\begin{align}
{H_T} = {H_S} + {H_E},\label{eq:H_T}
\end{align}
where
\begin{align}
{H_S} =& {\Delta _a}{a^\dag }a + {\Delta _{b_1}}b_1^\dag {b_1} + {\Delta _{b_2}}b_2^\dag {b_2} + {g_1}({a^\dag }{b_1} + ab_1^\dag ) \nonumber\\
 &+ {g_2}({a^\dag }{b_2} + ab_2^\dag ) + J(b_1^\dag {b_2} + {b_1}b_2^\dag ), \label{H_S}  \\
{H_E} =& \sum_k {{\Omega_k}} c_k^\dag {c_k} + i\sum_k {({C_k}ac_k^\dag  - C_k^*{a^\dag }{c_k})} \nonumber\\
 &+ \sum_s {{\varpi _s}} d_s^\dag {d_s} + i\sum_s {({D_s}ad_s^\dag  - D_s^*{a^\dag }{d_s})}, \label{H_E}
\end{align}
where $a$ (${a}^\dag$) denotes the annihilation (creation) operator of cavity $a$ with eigenfrequency $\omega_a$, while ${b}_1$ (${b}_1^\dag$) and ${b}_2$ (${b}_2^\dag$) denote those of cavities $b_1$ and $b_2$ with eigenfrequencies $\omega_{b_1}$ and $\omega_{b_2}$. Additionally, ${\upsilon _k}$ and ${{\mu _s}}$ correspond to the eigenfrequencies of the environmental modes ${c_k}$ and ${d_s}$, respectively. Within the rotating frame, the frequency detunings are denoted as ${\Delta _a} = {\omega _a} - {\omega _p}$, ${\Delta _{b_1}} = {\omega _{b_1}} - {\omega _p}$, ${\Delta _{b_2}} = {\omega _{b_2}} - {\omega _p}$, ${\Omega _k} = {\upsilon _k} - {\omega _p}$, and ${\varpi _s} = {\mu _s} - {\omega _p}$. 
Starting from the Heisenberg equations for the system and environment operators in Eq.~(\ref{eq:H_T}), and after eliminating the environmental degrees of freedom, we obtain the following non-Markovian integro-differential equations for the cavity modes (see Appendix~\ref{non-Markovian input–output} for more details)
\begin{equation}
\begin{aligned}
\dot a =  &- i{\Delta _a}a - i{g_1}{b_1} - i{g_2}{b_2} + K_1(t)  + K_2(t)\\
 &- \int_0^t {d\tau a} (\tau )\left[ {{f_1}(t - \tau ) + {f_2}(t - \tau )} \right],\\
{{\dot b}_1} =&  - i({\Delta _{{b_1}}} - i{\gamma _1}){b_1} - i{g_1}a - iJ{b_2},\\
{{\dot b}_2} =&  - i({\Delta _{{b_2}}} - i{\gamma _2}){b_2} - i{g_2}a - iJ{b_1}, \label{eq:system-dynamics}
\end{aligned}
\end{equation}
where the environmental correlation functions $f_{1}(t)$ and $f_{2}(t)$ are given by Eq.~(\ref{eq:memory kernel}), while the explicit forms of $K_{1}(t), K_{2}(t)$ can be found below Eq.~(\ref{eq:cavity-operator}). In this work, the non-Markovian environments are characterized by Lorentzian spectral densities, whose explicit forms are given in Eq.~(\ref{eq:Lorentzian}). We note that the system dynamics described through the Lorentzian spectral densities can be equivalently realized by the pseudomode theory, which provides a Markovian embedding of the structured reservoirs and is detailed in Appendix~\ref{Pseudomode analysis}. For simplicity, we consider the case where the two non-Markovian environments provided by the two ports are identical, i.e., the non-Markovian environmental spectral widths and the cavity dissipation rates coupled to the two ports respectively satisfy $\lambda_{1} = \lambda_{2}\equiv \lambda$ and $\Gamma_{1} = \Gamma_{2}\equiv \Gamma$. The general case of two different non-Markovian environments ($\lambda_{1} \neq \lambda_{2}$ and $\Gamma_{1} \neq \Gamma_{2}$) is discussed in Appendix~\ref{General case-double}.

\textit{The CPA condition.} For suitably chosen parameters, CPA occurs when the output fields vanish, i.e., $a_1^{(\rm out)}(t)=a_2^{(\rm out)}(t)=0$. The system exhibits an effective gain through the CPA of the two input fields at cavity $a$, allowing the Hamiltonian to be pseudo-Hermitian, which can be identified by analyzing the total output spectrum (see Sec.~\ref{section IIIA}).
To analyze the CPA condition, we perform a modified Laplace transform~\cite{Uchiyama1282009,Shen1222015} to Eq.~(\ref{eq:system-dynamics}) and obtain (derivation details can be found in Appendix~\ref{Derivation and discussion of Eq16})
\begin{align}
 &- i\omega a(\omega ) =  - \left[ {i{\Delta _a} + {f_1}(\omega ) + {f_2}(\omega )} \right]a(\omega ) - i{g_1}{b_1}(\omega )\nonumber\\
 &- i{g_2}{b_2}(\omega ) + {{\tilde \kappa }_1}(\omega )a_1^{({\rm{in}})}(\omega ) + {{\tilde \kappa }_2}(\omega )a_2^{({\rm{in}})}(\omega ),
\label{modified Laplace transform}\\&-i\omega b_1(\omega)=-i(\Delta_{b_1}-i\gamma_{1})b_1(\omega)-ig_1a(\omega)-iJb_2(\omega),\nonumber\\
&-i\omega b_2(\omega)=-i(\Delta_{b_2}-i\gamma_{2})b_2(\omega)-ig_2a(\omega)-iJb_1(\omega),\nonumber
\end{align}
where we have ${f_{1(2)}}(\omega ) = \int_0^\infty  {f_{1(2)}^*} (t'){e^{i\omega t'}}dt'$, ${{\tilde \kappa }_{1(2)}}(\omega ) = \int_{ - \infty }^0 {\kappa _{1(2)}^*} (t'){e^{i\omega t'}}dt'$, and $a_{1(2)}^{(\rm{in})}(\omega ) = \int_0^\infty  {a_{1(2)}^{(\rm{in})}} (t'){e^{i\omega t'}}dt'$. Then, the cavity field $a\left( \omega  \right)$ follows directly from Eq.~(\ref{modified Laplace transform}) and is given by 
\begin{align}
a{(\omega)}=\frac{\tilde{\kappa}_{1}(\omega )a_{1}^{(\rm{in})}(\omega)+\tilde{\kappa}_{2}(\omega )a_{2}^{(\rm{in})}(\omega)}{f_1(\omega)+f_2(\omega)+i(\Delta_{a}-\omega)+\sigma(\omega)},
\label{eq:solution cavity-a}
\end{align}
where $\sigma (\omega ) = (i{{{g^2_1}}{\Theta _2} + i{{{{g^2_2}}}{\Theta _1} + 2i{g_1}{g_2}J}})/({{{J^2} - {\Theta _1}{\Theta _2}}})$ denotes the self-energy of cavities $b_1$ and $b_2$ with ${\Theta _{1(2)}} = ( {{\omega _{{b_{1(2)}}}} - \omega } ) - i{\gamma _{1(2)}}$. Moreover, the non-Markovian input-output relation~(\ref{eq:input-output relation}) in the frequency domain reads
\begin{align}
    a_{1(2)}^{(\rm{in})}(\omega) +a_{1(2)}^{(\rm{out})}(\omega) =a(\omega)\kappa_{1(2)}(-\omega). \label{eq:input-output-frequency}
\end{align}
The CPA occurs when the following two conditions are satisfied simultaneously, as obtained by substituting Eq.~(\ref{eq:solution cavity-a}) into Eq.~(\ref{eq:input-output-frequency}) with $a_{1(2)}^{\left( {\rm{out}} \right)}\left( \omega_{\rm{CPA}}  \right) = 0$, then we have
\begin{align}
\frac{{a_1^{(\rm{in})}({\omega _{\rm{CPA}}})}}{{a_2^{(\rm{in})}({\omega _{\rm{CPA}}})}} = \frac{{{\kappa _1}( - {\omega _{\rm{CPA}}})}}{{{\kappa _2}( - {\omega _{\rm{CPA}}})}},\label{eq:CPA condition-1}
\end{align}
and
\begin{align}
{\omega _{\rm{CPA}}} &=  i\left[ {{{\tilde \kappa }_1}({\omega _{\rm{CPA}}}){\kappa _1}( - {\omega _{\rm{CPA}}}) + {{\tilde \kappa }_2}({\omega _{\rm{CPA}}}){\kappa _2}( - {\omega _{\rm{CPA}}})} \right]\nonumber\\
 &- i\left[ {{f_1}({\omega _{\rm{CPA}}}) + {f_2}({\omega _{\rm{CPA}}}) + \sigma \left( {{\omega _{\rm{CPA}}}} \right)} \right] + {\Delta _a},\label{eq:CPA condition-2}
\end{align}
where ${{\omega _{\rm{CPA}}}}$ represents the probe frequency at which CPA occurs, while ${\sigma \left( {{\omega _{\rm{CPA}}}} \right)}$ is obtained from $\sigma (\omega )$ below Eq.~(\ref{eq:solution cavity-a}). 
Conditions~(\ref{eq:CPA condition-1}) and~(\ref{eq:CPA condition-2}) indicate that the occurrence of CPA depends on both the tuning of system parameters and choosing appropriate input fields, where the two input fields must share the same phase and maintain an amplitude ratio of ${{{\kappa _1}( - {\omega _{\rm{CPA}}})} \mathord{\left/{\vphantom {{{\kappa _1}( - {\omega _{\rm{CPA}}})} {{\kappa _2}}}} \right.\kern-\nulldelimiterspace} {{\kappa _2}}}( - {\omega _{\rm{CPA}}})$, which can be realized experimentally by using a tunable phase shifter and a variable attenuator~\cite{Zhang3682017}.

\textit{Effective Hamiltonian and pseudo-Hermiticity}. In the CPA regime, Eq.~(\ref{eq:input-output relation}) reduces to
${a}_{1(2)}^{(\rm{in})}(t)=\int_{0}^{t}\kappa_{1{(2)}}(\tau-t){a}(\tau)d\tau.$ By taking this into account and defining $2gX(t): = -K_1 (t) - K_2 (t) + \int_0^t d\tau a (\tau )\left[ {{f_1}(t - \tau ) + {f_2}(t - \tau )} \right] $ with $g =\sqrt{{\lambda \Gamma}/{2}}$, Eq.~(\ref{eq:system-dynamics}) becomes ${\bf{\dot V}} =  - i{H_{\rm{eff}}}{\bf{V}}$, where ${\bf{V}} = {(a,{b_1},{b_2},X)^T}$ and the effective non-Markovian Hamiltonian $H_{\rm{eff}}$ can be expressed as
\begin{align}
{H}_{\rm{eff}}=\begin{pmatrix}\Delta_{a}&g_1&g_2&-2ig\\g_1&\Delta_{b_1}-i\gamma_{1}&J&0\\g_2&J&\Delta_{b_2}-i\gamma_{2}&0\\ig&0&0&i\lambda\end{pmatrix}.
\label{eq:effective Hamiltonian}
\end{align}

It is important to emphasize that the effective gain in $H_{\text{eff}}$ in Eq.~(\ref{eq:effective Hamiltonian}) does not correspond to physical gain in the open system. It emerges from the CPA, where the two input fields are precisely tailored in amplitude and phase so that they interfere destructively, and the output fields vanish identically. Under this specific boundary condition, the system behaves as if it has an effective gain, but this is not actual energy creation. In the unconditional description, the system remains purely dissipative, with no net gain. Thus, the non-Hermitian Hamiltonian $H_{\text{eff}}$ describes the conditional dynamics under CPA, not the intrinsic dynamics of the system.

Next, we verify under what parameter constraints the effective Hamiltonian~(\ref{eq:effective Hamiltonian}) satisfies pseudo-Hermiticity. Hamiltonian~(\ref{eq:effective Hamiltonian}) has a quartet of eigenvalues. According to the method in Ref.~\cite{Mostafazadeh2052002}, $H_{\rm{eff}}$ is pseudo-Hermitian if all four eigenvalues are real. From the spectral properties of pseudo-Hermitian Hamiltonians, this condition is equivalent to demanding that ${\rm{det}}({H_{{\rm{eff}}}} - { E}{\mathbb{I}}) = 0$ and its complex conjugate $\mathrm{det}({H}_{\rm{eff}}^{*}- { E}{\mathbb{I}})=0$ possess identical solutions, where $\mathbb{I}$ denotes the identity matrix and ${ E}$ denotes the eigenvalues of Eq.~(\ref{eq:effective Hamiltonian}). Expanding the characteristic equation and its complex conjugate, and then comparing the corresponding coefficients, yield a set of constraints under which the effective Hamiltonian is pseudo-Hermitian as follows
\begin{align}
&{({g_1})^2}{\gamma _2} + {({g_2})^2}{\gamma _1} - \lambda ({\Delta _{{b_1}}}{\Delta _{{b_2}}} + \lambda \Gamma  - {\gamma _1}{\gamma _2} - {J^2}) = 0,\nonumber\\
&{({g_2})^2}{\Delta _{{b_1}}} + {({g_1})^2}{\Delta _{{b_2}}} + \left( {{J^2} + {\gamma _1}{\gamma _2} - {\Delta _{{b_1}}}{\Delta _{{b_2}}}} \right){\Delta _a}\nonumber\\
 &- 2J{g_1}{g_2} - \Gamma ({\gamma _2}{\Delta _{{b_1}}} + {\gamma _1}{\Delta _{{b_2}}}) = 0,\label{eq:pseudo-Hermitian}\\
&\lambda  - {\gamma _1} - {\gamma _2} = 0,\;{\gamma _1}{\Delta _{{b_1}}} + {\gamma _2}{\Delta _{{b_2}}} = 0.\nonumber
\end{align}

\section{Higher-order exceptional points} \label{section III}
In this section, we first present numerical results for the example discussed in Sec.~\ref{section II}, demonstrating that when the effective Hamiltonian satisfies pseudo-Hermiticity, the non-Markovian structured reservoir—compared with its Markovian counterpart—can induce a higher-order EP, namely EP4. We further show that EP4 can be tracked and identified from the spectral structure of the total output spectrum. Moving beyond this case, we discuss the generic non-Hermitian scenario. Finally, we classify the possible higher-order EPs in the system and expand the discussion of topological invariants associated with higher-order EPs.

\subsection{Emergence of higher-order EPs} \label{section IIIA}

\begin{figure}[b]
    \centering
    \includegraphics[width=0.486 \textwidth]{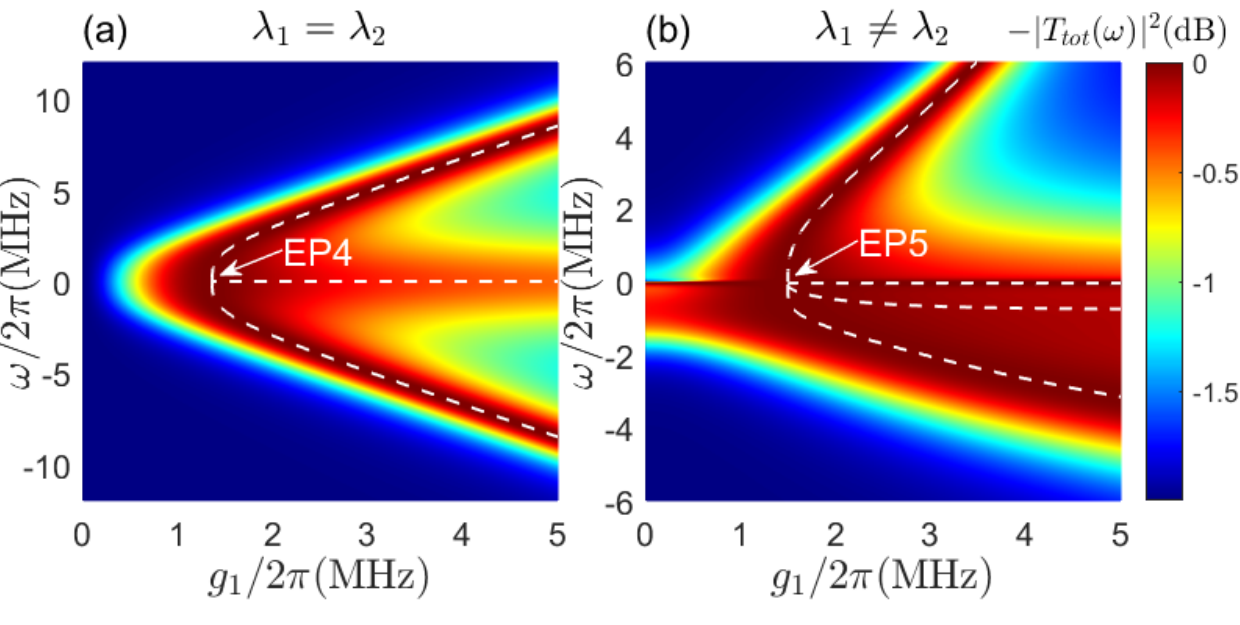} 
    \caption{The total output spectrum~(\ref{eq:output-total}) varies with the coupling strength $g_1$ and the probe frequency $\omega$  for (a) equal (corresponding to Eqs. (\ref{eq:pseudo-Hermitian}) and (\ref{eq:characteristic-f})) and (b) unequal (corresponding to Eqs. (\ref{x5}) and (\ref{constraint})) cases of environmental spectral widths. The white dashed lines and the dark red contours correspond to the real parts of the eigenvalues and the CPA condition, respectively, thereby demonstrating the emergence of EPs. The parameters chosen are $\gamma_1/2\pi=\gamma_2/2\pi=1.885$ MHz, $\lambda_1/2\pi=\lambda_2/2\pi=3.77{\rm{MHz}}$, $g_2=g_1$, $J=0$, ${\Delta _{{b_2}}}/2\pi =  - {\Delta _{{b_1}}}/2\pi=-0.866$ MHz, $\Gamma_1 = \Gamma_2 =(g_1^2+\gamma_2^2+\Delta_{b_1}^2)/(2\gamma_2)$, and ${\Delta _a} = 0$ for (a); ${{{\gamma _1}} \mathord{\left/
 {\vphantom {{{\gamma _1}} {2\pi }}} \right.
 \kern-\nulldelimiterspace} {2\pi }} = 2$ MHz, ${{{\gamma _2}} \mathord{\left/
 {\vphantom {{{\gamma _2}} 2\pi }} \right.
 \kern-\nulldelimiterspace} 2\pi } = 1$ MHz, $\lambda_1/2\pi=0.079 {\rm{MHz}}$, $\lambda_2/2\pi=2.921 {\rm{MHz}}$, $g_2/2\pi=0.589$MHz, $J/2\pi=0.0058 {\rm{MHz}}$,  and $\;{\Delta_{{b_2}}}/2\pi = \; - 2{\Delta _{{b_1}}}/2\pi =-0.787 {\rm{MHz}}$ for (b), while the remaining parameters $\Delta_a$, $\Gamma_1$, $\Gamma_2$ are given by Eq.~(\ref{constraint})}.
    \label{fig:Fig2}
\end{figure}

\begin{figure}[t]
   \centerline{
   \includegraphics[width=8.5cm, height=8.0cm, clip]{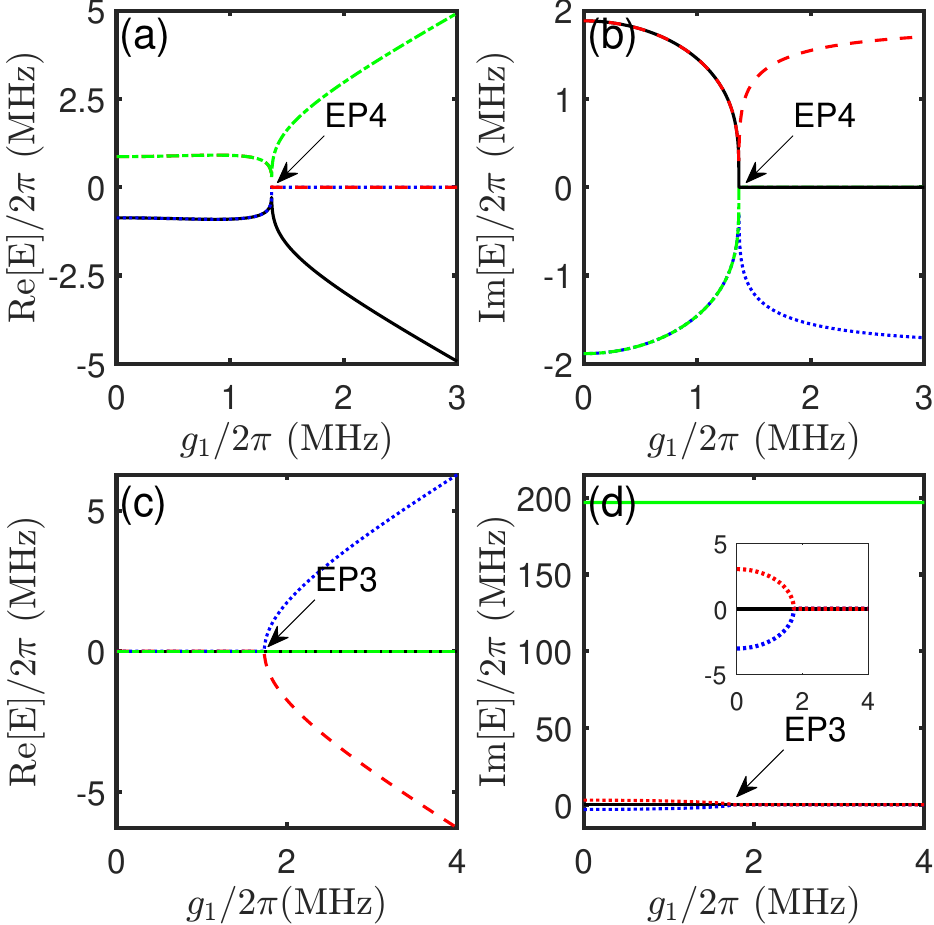}}
   \caption{The real and imaginary parts of the eigenvalues obtained from the characteristic equation~(\ref{eq:characteristic-f}) with pseudo-Hermitian condition (\ref{eq:pseudo-Hermitian}) as functions of the coupling strength $g_1$. (a) and (b) correspond to the small environmental spectral widths $\lambda_1/2\pi=\lambda_2/2\pi=3.77{\rm{MHz}}$ as in Fig.~\ref{fig:Fig2}(a), whereas (c) and (d) correspond to very large environmental spectral widths $\lambda_1/2\pi=\lambda_2/2\pi=200{\rm{MHz}}$ (Markovian approximation). The parameters chosen are $\gamma_1/2\pi=\gamma_2/2\pi=1.5{\rm{MHz}}$, $\Gamma_1/2\pi=\Gamma_2/2\pi=3{\rm{MHz}}$, $J=0$, $\Delta_a=0$, $g_2=g_1$, and $\Delta_{b_2}=-\Delta_{b_1}$.}\label{fig:Fig3}
 \end{figure}

We continue with the case in Sec.~\ref{section II}, where the two ports of cavity $a$ are coupled to two non-Markovian structured reservoirs. By deriving the total output spectrum of cavity $a$ coupled to two non-Markovian structured reservoirs, we numerically demonstrate the pseudo-Hermiticity of the system and further show that higher-order EPs can be read out through spectral analysis.
Under condition~(\ref{eq:CPA condition-1}), Eq.~(\ref{eq:input-output-frequency}) can be rewritten in the form $a_{1\left( 2 \right)}^{\left( {\rm{out}} \right)}\left( \omega  \right) = {T_{1\left( 2 \right)}}\left( \omega  \right)a_{1\left( 2 \right)}^{\left( {\rm{in}} \right)}\left( \omega  \right),$
where ${T_{1\left( 2 \right)}}\left( \omega  \right)$ is the scattering coefficient at port 1 or 2, and its corresponding scattering probability is given by ${| {{T_{1(2)}}\left( \omega  \right)} |^2}$. Explicitly, $T_1$ and $T_2$ are expressed as
\begin{align}
T_1(\omega)=\frac{\kappa_{1}(-\omega)\tilde\kappa_{1}(\omega)+\kappa_{2}(-\omega)\tilde\kappa_{2}(\omega)}{f_{1}(\omega)+f_{2}(\omega)+i(\Delta_{a}-\omega)+\sigma(\omega)}-1, \label{eq:explicit-T}
\end{align}
and ${T_2}\left( \omega  \right) = {T_1}\left( \omega  \right)$. To further describe the input-output characteristics of the whole system, we introduce the total output spectral function of cavity $a$, denoted as
\begin{align}
    {\left| {{T_{\rm{tot}}}\left( \omega  \right)} \right|^2} = {\left| {{T_1}\left( \omega  \right)} \right|^2} + {\left| {{T_2}\left( \omega  \right)} \right|^2} = 2{\left| {{T_1}\left( \omega  \right)} \right|^2}.\label{eq:output-total}
\end{align}
We note that once the remaining necessary condition~({\ref{eq:CPA condition-2}}) for CPA is also satisfied, the total output spectrum ${\left| {{T_{{\rm{tot}}}}\left( {{\omega _{{\rm{CPA}}}}} \right)} \right|^2}$ becomes zero.

We now numerically observe the emergence of higher-order EPs through the total output spectrum under the pseudo-Hermitian condition. In the simulations, we examine two cases distinguished by   identical environments ($\lambda_{1} = \lambda_{2}\equiv \lambda$ and $\Gamma_{1} = \Gamma_{2}\equiv \Gamma$, see Eq.~(\ref{eq:effective Hamiltonian})); different environments ($\lambda_{1} \neq \lambda_{2}$ and $\Gamma_{1} \neq \Gamma_{2}$, see Appendix~\ref{General case-double}). With this, we observe the emergence of EP4 and EP5 induced by non-Markovian reservoirs. In Fig.~\ref{fig:Fig2}(a) and (b), we present the total output spectrum as a function of the coupling strength $g_1$ and probe frequency $\omega$ for the equal and unequal cases of environmental spectral widths, respectively. The minima in the total output spectrum, highlighted by the dark red contour, correspond to the condition of CPA. They reveal that the CPA frequency ${\omega _{\rm{CPA}}}$ agrees well with the real parts of the eigenvalues of the effective Hamiltonian~(\ref{eq:effective Hamiltonian}) under the pseudo-Hermiticity condition~(\ref{eq:pseudo-Hermitian}), which are marked by the white dashed lines. The above analysis not only demonstrates the experimental potential for probing the energy spectrum structure by monitoring the total output spectrum, but also establishes the feasibility of using this approach to track and identify critical singular points such as EP4 in the non-Markovian system. 
Furthermore, we determine EP4 by solving the characteristic equation of the effective Hamiltonian~(\ref{eq:effective Hamiltonian}) under the pseudo-Hermitian condition~(\ref{eq:pseudo-Hermitian}) as follows
\begin{align}
{x^4} + a{x^3} + b{x^2} + cx + d = 0,\label{eq:characteristic-f}
\end{align}
where $x = {\rm E} - {\Delta _a}$, and ${\rm E}$ is the eigenvalue of Eq.~(\ref{eq:effective Hamiltonian}) under the pseudo-Hermiticity condition~(\ref{eq:pseudo-Hermitian}). The coefficients in Eq.~(\ref{eq:characteristic-f}) are explicitly given by $a =  - {\Delta _{{b_1}}} - {\Delta _{{b_2}}} + 3{\Delta _a}$, $b =  - g_1^2 - g_2^2 - {J^2} - \Gamma \lambda  + {\lambda ^2} - {\gamma _1}{\gamma _2} + {\Delta _{{b_1}}}{\Delta _{{b_2}}} + 3\Delta _a^2 - 2({\Delta _{{b_1}}} + {\Delta _{{b_2}}}){\Delta _a}$, $c = 2g_2^2({\Delta _{{b_1}}} - {\Delta _a}) + 2g_1^2({\Delta _{{b_2}}} - {\Delta _a}) - 4J{g_1}{g_2} - \lambda ({\gamma _2}{\Delta _{{b_1}}} + {\gamma _1}{\Delta _{{b_2}}}) + \lambda (\lambda  - 2\Gamma ){\Delta _a} - ({\Delta _{{b_1}}} + {\Delta _{{b_2}}})\Delta _a^2 + \Delta _a^3$, and $d = 2\Gamma ({\gamma _2}{\Delta _{{b_1}}} + {\gamma _1}{\Delta _{{b_2}}}){\Delta _a} - ({J^2} + \Gamma \lambda  + {\gamma _1}{\gamma _2} - {\Delta _{{b_1}}}{\Delta _{{b_2}}})\Delta _a^2 - g_2^2(\lambda {\gamma _1} + \Gamma {\lambda _2} + \Delta _a^2) - g_1^2(\lambda {\gamma _2} + \Gamma {\lambda _1} + \Delta _a^2) + {\Gamma ^2}{\lambda ^2}$. 
According to the discriminant of a quartic equation, when the condition $B^2-4AC=0$ is satisfied, the equation possesses four degenerate real roots, where $A=D^2-3F$, $B=DF-9{\bar E}^2$, and $C=F^2-3D{\bar E}^2$ with $D = 3{a^2} - 8b, {\bar E} =  - {a^3} + 4ab - 8c,~F = 3{a^4} + 16{b^2} - 16{a^2}b + 16ac - 64d$. 
When $D={\bar E}=F=0$, Eq.~(\ref{eq:characteristic-f}) possesses a quadruple real root, i.e., ${{\rm E}_1} = {{\rm E}_2} = {{\rm E}_3} = {{\rm E}_4} =  - {a}/{4} =  - {{2b}}/{{3a}} =  - {{3c}}/{{2b}} =  - {{4b}}/{c}$. 
Notably, the above analysis is restricted to the symmetry case with ${\gamma_1} = {\gamma_2}$. For the more general situation of broken parameter symmetry, i.e., ${\gamma _1} \ne {\gamma _2}$, detailed analyses can be found in Appendix~\ref{Asymmetry analysis}.
 
\begin{figure}[t]
   \centerline{
   \includegraphics[width=8.5cm, height=8.cm, clip]{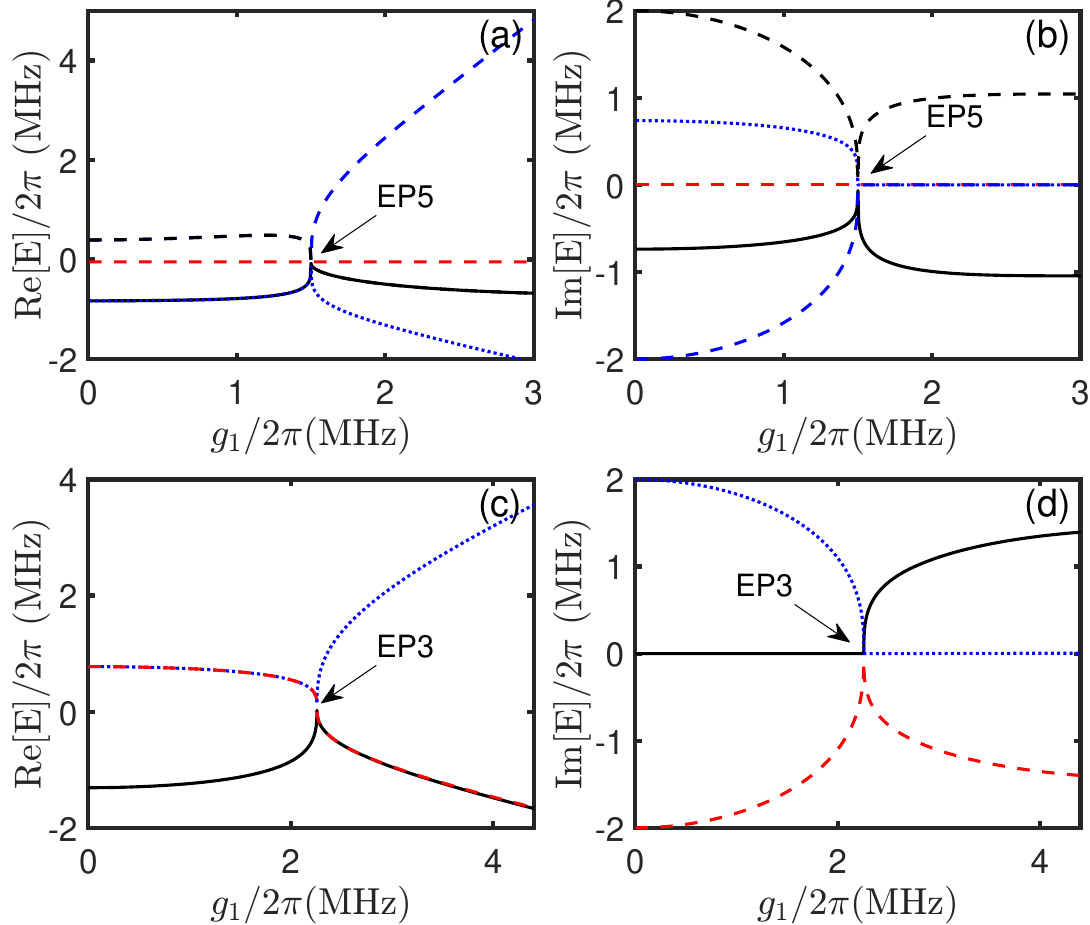}}\caption{With Eq. (\ref{constraint}), the conversion of eigenvalues from the non-Markovian to the Markovian regime at the EPs is analyzed as a function of the coupling strength $g_1$. (a) and (b) respectively plot the real and imaginary parts of the eigenvalues in Eq. (\ref{x5}) of the effective Hamiltonian ${H}_{5}$ in the non-Markovian case with $\lambda_1/2\pi = 0.079$ MHz and $\lambda_2 /2\pi= 2.921$ MHz.  Other parameters are chosen as same as those in Fig.~\ref{fig:Fig2}(b). (c) and (d) present the real and imaginary parts under the Markovian approximation. The parameters chosen are $\gamma_1/2\pi=2$MHz, $\gamma_2/2\pi=1$MHz, $\Delta_{b_2}/2\pi=-0.787$MHz,
  $\Delta_{b_1}/2\pi=0.777$MHz,
  $J/2\pi=0.0058$MHz,
  $\Delta_a/2\pi=0.256$MHz, ${g_2}{\gamma _1} = \sqrt {{\gamma _2}({\gamma _1} + {\gamma _2}){{({\Delta _{{b_2}}} - {\Delta _a})}^2} + {\gamma _1}({\gamma _1} + {\gamma _2})({\gamma _1}{\gamma _2} + {J^2}) - {\gamma _1}{\gamma _2}g_1^2}$, $\Gamma_2/2\pi=5$MHz, and $\Gamma_1/2\pi=1$MHz.}
  \label{fig:Fig14}
\end{figure}

Next, corresponding to Fig.~\ref{fig:Fig2}, we present the real and imaginary parts of the eigenvalues of the system as functions of the coupling strength $g_1$, and further demonstrate the increase in the EP order caused by difference in the environmental spectral widths $\lambda_1$ and $\lambda_2$. 
For the equal case ${\lambda _1} = {\lambda _2}$, Fig.~\ref{fig:Fig3}(a) and (b) exhibit that the critical coupling strength at EP4 is $g_{\mathrm{1EP4}}=1.367$ MHz. At this point, all four eigenvalues corresponding to Eq.~(\ref{eq:effective Hamiltonian}) coalesce into a single real value. Furthermore, for $g_1<g_{\mathrm{1EP4}}$, the four eigenvalues appear as two complex-conjugate pairs, whereas for $g_1>g_{\mathrm{1EP4}}$, the spectrum splits into two real eigenvalues and one complex-conjugate pair. For this case, we find that as the spectral width of the non-Markovian environments increases from being exceedingly narrow in Fig.~\ref{fig:Fig3}(a) and (b) to being remarkably broad in Fig.~\ref{fig:Fig3}(c) and (d), the non-Markovian reservoir effectively reduces to a Markovian one, and EP4 degenerates into EP3.


In Fig.~\ref{fig:Fig14}, we consider the second case where the environmental spectral widths are different ($\lambda_1 \neq \lambda_2$) with $\lambda_1 = 0.079$MHz and $\lambda_2 = 2.921$ MHz. Figure~\ref{fig:Fig14}(a) and~(b) respectively show the real and imaginary parts of the eigenvalues corresponding to Eq.~(\ref{5*5NM}) as functions of $g_1$ in the non-Markovian case with the critical coupling strength ${g}_{\text{1EP5}} = 1.5$ MHz at EP. When $g_1 < {g}_{\text{1EP5}}$, the five eigenvalues manifest as one real and two pairs of complex conjugates. At $g_1 = {g}_{\text{1EP5}}$ (i.e., at EP5), the five eigenvalues coalesce into a real number. With the parameter $g_2$ fixed at 0.589 MHz, the eigenvalues evolve as follows: when $g_1 > {g}_{\text{1EP5}}$, they appear as three real values and one pair of complex conjugates. The cases of Markovian EPs are presented in Fig.~\ref{fig:Fig14}(c) and~(d). In this case, the critical coupling strength is ${g}_{\text{1EP3}} /2\pi = 2.258$ MHz. When $g_1 < {g}_{\text{1EP3}}$, three eigenvalues are observed as one real and one pair of complex conjugates. At $g_1 = {g}_{\text{1EP3}}$, i.e., at EP3, three eigenvalues coalesce into a real number. When $g_1 > {g}_{\text{1EP3}}$, the eigenvalues appear as another real and one pair of complex conjugates. Consequently, we can observe the conversion from a fifth-order EP in a non-Markovian pseudo-Hermitian system to a third-order EP in a Markovian system.

\begin{figure}[t]
   \centerline{
   \includegraphics[width=8.5cm, height=8cm, clip]{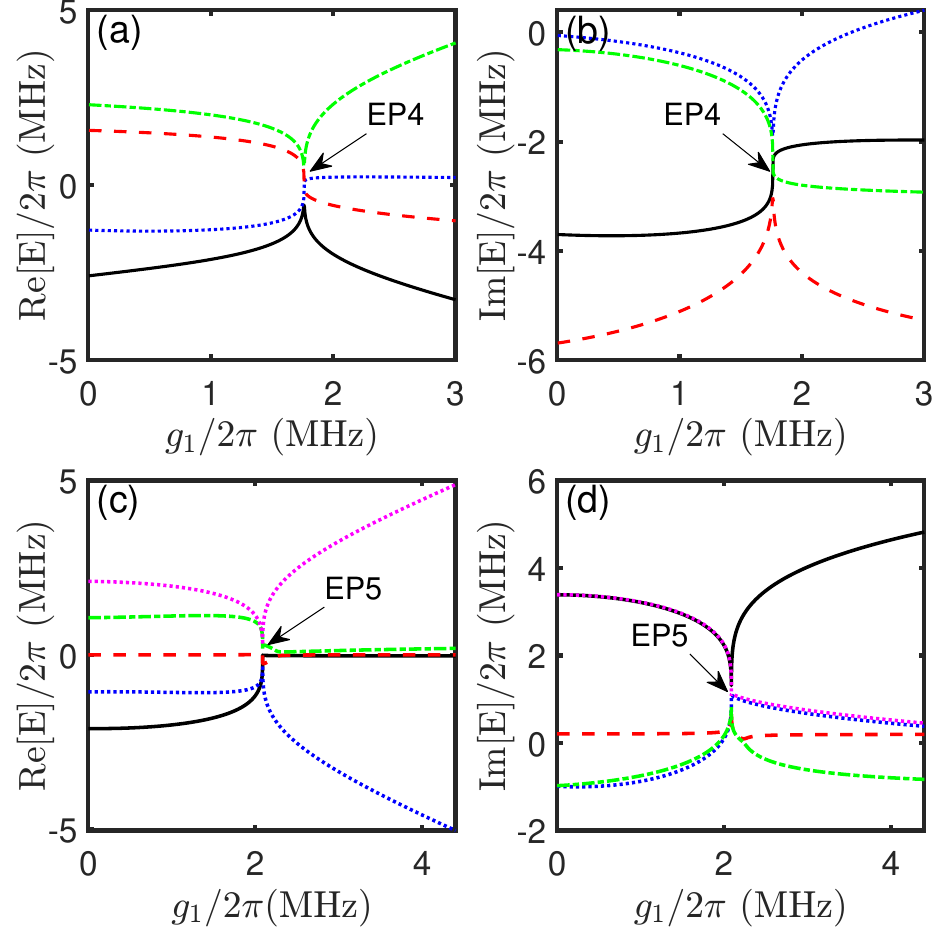}}
       \caption{The real and imaginary parts of the eigenvalues are plotted as functions of the coupling strength $g_1$ without pseudo-Hermitian condition. (a) and (b) present the case of equal environmental spectral width in Eq. (\ref{eq:characteristic-f}) by removing Eq. (\ref{eq:pseudo-Hermitian}), where ${{{\lambda _1}} \mathord{\left/
 {\vphantom {{{\lambda _1}} {2\pi }}} \right.
 \kern-\nulldelimiterspace} {2\pi }} = {{{\lambda _2}} \mathord{\left/
 {\vphantom {{{\lambda _2}} {2\pi }}} \right.
 \kern-\nulldelimiterspace} {2\pi }} = 2$MHz, ${\gamma _1}/2\pi = {\gamma _2}/2\pi=5.88$MHz, $g_2/2\pi=3.59$MHz, $J/2\pi=1$MHz, $\Delta_a=0$, $\Delta_{b_1}/2\pi=-\Delta_{b_2}/2\pi=1.29$MHz, and $\Gamma_2/2\pi=\Gamma_1/2\pi=3$MHz. The unequal case in Eq. (\ref{x5}) by removing Eq. (\ref{constraint}) is illustrated in~(c) and (d), where  ${{{\lambda _1}} \mathord{\left/
 {\vphantom {{{\lambda _1}} {2\pi }}} \right.
 \kern-\nulldelimiterspace} {2\pi }} = 0.2$MHz, ${{{\lambda _2}} \mathord{\left/
 {\vphantom {{{\lambda _2}} {2\pi }}} \right.
 \kern-\nulldelimiterspace} {2\pi }} = 6.83$MHz, ${\Gamma _2}/2\pi = 4.74$MHz, ${\gamma _1}/2\pi={\gamma _2}/2\pi=1$MHz, $\Gamma_1/2\pi=0.2$MHz, $g_2/2\pi=0.352$MHz, $J/2\pi=1$MHz, $\Delta_a=0$, and $\Delta_{b_1}/2\pi=-\Delta_{b_2}/2\pi=-0.347$MHz.}\label{fig:Fig4}
 \end{figure}
 
Figure~\ref{fig:Fig4} presents higher-order EPs in the generic non-Hermitian system by lifting pseudo-Hermitian constraint. The analysis of this process follows a method similar to that described for the pseudo‑Hermitian cases, and the corresponding characteristic equations can be directly obtained by removing the pseudo‑Hermitian conditions in Eq.~(\ref{constraint}). The four lines in Fig.~\ref{fig:Fig4}(a) and (b) represent the variations of the real parts and imaginary parts of four eigenvalues with respect to the coupling strength $g_1$, where the environmental spectral widths are equal ($\lambda_1=\lambda_2$), while Fig.~\ref{fig:Fig4}(c) and (d) show the corresponding variations of the real and imaginary parts of five eigenvalues with respect to the coupling strength $g_1$ for $\lambda_1 \neq \lambda_2$. Specifically, Fig.~\ref{fig:Fig4}(a) and (b) describe a non-Hermitian EP4 (${g}_{\text{1EP4}} /2\pi = 1.765$ MHz) in the non-Markovian regime corresponding to pseudo-Hermitian Fig.~\ref{fig:Fig3}(a) and (b). Figure~\ref{fig:Fig4}(c) and (d) denote a non-Hermitian EP5 (${g}_{\text{1EP5}} /2\pi = 2.0875$ MHz) in a non-Markovian system corresponding to pseudo-Hermitian Fig.~\ref{fig:Fig14}(a) and (b). We find that the EP4 and EP5 in non-Hermitian systems in Fig.~\ref{fig:Fig4} are different from those of pseudo-Hermitian systems in Figs.~\ref{fig:Fig3} and \ref{fig:Fig14}, which arises from the constraints imposed by pseudo-Hermitian conditions. 

\subsection{Classification of higher-order EPs} \label{section IIIB}

\begin{table}[t]
\caption{Symbols and their physical meanings.}\label{T:table I}
\begin{ruledtabular}
\begin{tabular}{ll}
\textrm{Symbol} & \textrm{Meaning} \\
\colrule
$g_{1(2)}$ & Coupling strength between cavity modes $a$ and $b_{1(2)}$ \\
$J$ & Linear coupling strength between cavities $b_{1}$ and $b_{2}$ \\
${\Gamma _{1(2)}}$ & Dissipation rate of cavity $a$ at port $1(2)$ \\
${{\lambda _{1(2)}}}$ & Spectral width of the non-Markovian bath at port $1(2)$\\
${\gamma _{1(2)}}$ & Markovian dissipation rate of cavity $b_{1(2)}$\\
${\Upsilon _{1\left( 2 \right)}}$ & Non-Markovian bath dissipation rate of cavity $b_{1(2)}$ \\
${\Lambda _{1\left( 2 \right)}}$ & Spectral width of the non-Markovian bath of cavity $b_{1(2)}$\\
\end{tabular}
\end{ruledtabular}
\end{table}

In the preceding discussion, we have shown that non-Markovian environments, in contrast to their Markovian counterparts, can induce higher-order EPs. However, the above analysis has focused on the simplest case, where only cavity $a$ is coupled to non-Markovian environments. The two structured non-Markovian reservoirs provided by the two ports are identical ($\lambda_{1} = \lambda_{2}\equiv \lambda$ and $\Gamma_{1} = \Gamma_{2}\equiv \Gamma$) or different (${\lambda _1} \ne {\lambda _2}$ and ${\Gamma _1} \ne {\Gamma _2}$). While the analysis of the simplest case is meaningful and provides us with a clear perspective on how non-Markovian structured reservoirs induce higher-order EPs, beyond this example, we now carry out a multi-parameter exploration of the higher-order EPs that may emerge in three-mode open optical systems. In this subsection, we present only the classification results of higher-order EPs based on different scenarios and parameter conditions, while the related details and derivation can be found in Appendices~\ref{General case-double}, \ref{Asymmetry analysis}, and ~\ref{Classification-details}. We emphasize that classifying higher-order EPs hinges on the memory effect, which scales inversely with environmental spectral width, and on the symmetric structure of the parameter space of the effective Hamiltonian. 

To facilitate the subsequent classification, Table~\ref{T:table I} lists the symbols involved in our classification of higher-order EPs together with a review of their corresponding physical meanings. In what follows, the analysis advances from the simplest to the most general case, in order of increasing parametric complexity, thereby expanding the effective dimensionality of a non-Markovian system.

First, as the simplest case, three optical cavities are coupled to four Markovian reservoirs with the parameter space satisfying the symmetry, i.e., ${\Gamma _1} = {\Gamma _2}$, ${\lambda _1} = {\lambda _2} \to \infty $ (when the environmental spectral widths are broad, e.g., ${\lambda _1} = {\lambda _2} = 200{\rm{MHz}}$ in practice, the memory effects can be neglected), ${\gamma _1} = {\gamma _2}$, ${g_1} = {g_2}$, and $J=0$; its effective Hamiltonian exhibits a symmetric 3$ \times $3 structure, whose eigenspectrum only allows the emergence of EP3. Moreover, once the symmetric structure is broken, e.g., ${\gamma _1} \ne {\gamma _2}$, ${g_1} \ne {g_2}$, and $J \ne 0$, the eigenspectrum undergoes asymmetric splitting, giving rise to EP2 in addition to EP3 shown in Appendix~\ref{Asymmetry analysis}.

Second, we consider the case of the cavity $a$ coupling to two identical non-Markovian structured reservoirs, with ${\Gamma _1} = {\Gamma _2}$ and ${\lambda _1} = {\lambda _2} = 3.77\,\mathrm{MHz}$, where narrow environmental spectral widths lead to strong non-Markovian memory effects. The cavities $b_1$ and $b_2$ are coupled to Markovian reservoirs, with ${\gamma _1} = {\gamma _2}$, while the remaining parameters satisfy ${g_1} = {g_2}$ and $J = 0$.
This yields a $4\times4$ effective Hamiltonian, thus allowing the emergence of EP4. Furthermore, we consider the case of system coupling to two non-Markovian environments. To be specific, cavity $a$ is coupled to two distinct non-Markovian structured reservoirs, with ${\Gamma _1} \ne {\Gamma _2}$ and ${\lambda _1} \ne {\lambda _2}$ (each corresponding to a narrow environmental spectral width, e.g., ${\lambda _1} = 0.079\,\mathrm{MHz}$ and ${\lambda _2} = 2.921\,\mathrm{MHz}$). The cavities $b_1$ and $b_2$ are coupled to two Markovian reservoirs, with ${\gamma _1} \ne {\gamma _2}$, while all other parameters are chosen appropriately. As a result, the system is governed by a 5$ \times $5 effective Hamiltonian and thereby enables the emergence of EP5. The relevant details are provided in Appendix~\ref{General case-double}.

Third, under the premise that cavity $a$ is coupled to two distinct non-Markovian structured reservoirs,
i.e., ${\Gamma _1} \ne {\Gamma _2}$ and ${\lambda _1} \ne {\lambda _2}$, we further consider the case where cavities $b_1$ and $b_2$ are each coupled to the same non-Markovian reservoir, satisfying ${\Upsilon _1} = {\Upsilon _2}$ and ${\Lambda _1} = {\Lambda _2}$, while the other parameters are chosen appropriately. In this case, the analysis extends beyond pseudo-Hermiticity, as the system is coupled to three distinguishable non-Markovian environments, resulting in a $6\times6$ effective Hamiltonian that enables the emergence of EP6. Lastly, we extend our analysis to the most general case, where all three optical cavities are coupled to distinguishable non-Markovian environments.
Specifically, cavity $a$ is coupled to two distinct non-Markovian structured reservoirs, with ${\Gamma _1} \ne {\Gamma _2}$ and ${\lambda _1} \ne {\lambda _2}$, while cavities $b_1$ and $b_2$ are each coupled to a different non-Markovian structured reservoir, with ${\Upsilon _1} \ne {\Upsilon _2}$ and ${\Lambda _1} \ne {\Lambda _2}$. All other parameters are chosen appropriately.
Accordingly, the system is effectively coupled to four distinct non-Markovian environments, yielding a $7\times7$ effective Hamiltonian and allowing the emergence of EP7 in Appendix~\ref{Classification-details}.

Furthermore, we point out that EPs are not only algebraic branch singularities of non-Hermitian spectra, but also topological defects in parameter space~\cite{Dembowski787,Ding0072016,Bergholtz0052021}. For a conventional EP2, the topology is usually characterized by the eigenvalue vorticity associated with the square-root branch point. For higher-order EPs of order \(k>2\), however, pairwise eigenvalue winding alone is generally insufficient to distinguish a genuine multifold EP from several lower-order EPs enclosed by the same loop. A more appropriate characterization is provided by the homotopy invariant associated with the resultant vector of the characteristic polynomial~\cite{Yoshida186602,Yoshida012021}. In particular, for a symmetry-protected \(k\)th-order EP, where \(k\) denotes the algebraic multiplicity of the exceptional degeneracy, i.e., the number of coalescing eigenvalues and eigenstates, the EP has codimension \(k-1\). Therefore, in the transverse parameter space, it can be enclosed by a closed \((k-2)\)-dimensional surface \(\Sigma_{k-2}\simeq S^{k-2}\), where \(S^{k-2}\) denotes a \((k-2)\)-dimensional sphere. The normalized resultant vector then defines a map \(\hat{\mathbf{R}}:\Sigma_{k-2}\to S^{k-2}\), and the degree of this map gives the corresponding topological charge.

This general construction applies directly to the EP4 discussed in Sec.~\ref{section IIIA}. For \(k=4\), the symmetry-protected codimension is \(k-1=3\), so the EP4 can be enclosed by a closed two-dimensional surface \(\Sigma_2\simeq S^2\) in the pseudo-Hermitian parameter submanifold. Under the pseudo-Hermiticity constraint, the quartic characteristic equation in Eq.~(\ref{eq:characteristic-f}) has real coefficients, and the quadruple-root condition associated with the EP4 can be expressed by the simultaneous vanishing of the degeneracy vector
\(\mathbf{R}_4=(D,\bar E,F)\),
where \(D\), \(\bar E\), and \(F\) are defined below Eq.~(\ref{eq:characteristic-f}). This three-component degeneracy vector plays the role of the resultant vector for the EP4. Provided that \(|\mathbf{R}_4|\neq 0\) on the enclosing surface \(\Sigma_2\), the normalized vector
\(\hat{\mathbf{R}}_4=\mathbf{R}_4/|\mathbf{R}_4|\)
defines a map
\(\hat{\mathbf{R}}_4:\Sigma_2\to S^2\).
The integer degree of this map characterizes the topological charge of the EP4~\cite{Yoshida186602,Yoshida012021}. Explicitly, this charge can be written as
\begin{align}
Q_{4}
=
\frac{1}{4\pi}
\int_{\Sigma_{2}}
\hat{\mathbf{R}}_{4}\cdot
\left(
\partial_{u}\hat{\mathbf{R}}_{4}
\times
\partial_{v}\hat{\mathbf{R}}_{4}
\right)
\,du\,dv,
\end{align}
where \(u\) and \(v\) parameterize the enclosing surface \(\Sigma_2\). The integer \(Q_4\) is invariant under smooth deformations of \(\Sigma_2\), as long as the surface does not pass through another degeneracy. Similarly, for the EP5 described by the quintic equation in Appendix~\ref{General case-double}, the quintuple-root condition \(L' = M' = N' = P' = 0\) defines a four-component degeneracy vector \(\mathbf{R}_5 = (L', M', N', P')\), whose normalized map from a closed three-dimensional hypersurface to \(S^3\) gives the corresponding higher-dimensional resultant winding number~\cite{Yoshida012021}. Thus, non-Markovianity in our work not only increases the algebraic order of EPs, but also enlarges the dimensionality of the associated topological map, allowing the system to support higher-order exceptional defects with quantized resultant winding charges. Additionally, we note that recent superconducting-circuit experiments have measured the topological invariant associated with an EP3 in quantum three-mode systems~\cite{Han02839}.

The above analysis supports a common intuition: each distinguishable non-Markovian structured reservoir with a Lorentzian spectrum increases the dimension of the effective Hamiltonian by one. This conclusion is consistent with pseudomode theory, where coupling an optical mode to a non-Markovian reservoir is equivalently mapped to its coupling with an auxiliary mode that in turn interacts with a Markovian reservoir, thereby reflecting the enlargement of the effective dynamical space induced by memory effects. Physically, the memory effect rooted in non-Markovian environments scales inversely with the environmental spectral width, becoming significant for narrow widths and manifesting as an enlargement of the effective Hamiltonian, thereby giving rise to higher-order EPs. In other words, the introduction of memory effects increases the complexity of the parameter space of the system, thereby inducing the emergence of higher-order EPs. In the non-Markovian limit (corresponding to the Markovian approximation) $\lambda \to \infty$ the memory effects become local, the auxiliary modes decouple, and the EP order reduces.

\section{Possible experimental implementation}\label{experimental analysis}

We now discuss the experimental feasibility of observing the proposed non-Markovian-induced higher-order EPs in superconducting circuits. The three optical modes \(a\), \(b_{1}\), and \(b_{2}\) in our theoretical model can be implemented by three microwave resonator modes, such as coplanar-waveguide or lumped-element superconducting resonators. Their bare resonance frequencies can be chosen in the range of \(4{-}8~\mathrm{GHz}\), while all relevant dynamics in the rotating frame are controlled by MHz-scale detunings, couplings, and engineered linewidths. Such parameter scales are natural in circuit QED, where microwave resonators, superconducting artificial atoms, tunable couplers, and engineered electromagnetic environments have been widely developed~\cite{Xiang6232013,Gu12017,Blais025005,Krantz021318}. The detunings \(\Delta_{a}\), \(\Delta_{b_{1}}\), and \(\Delta_{b_{2}}\) can be adjusted by SQUID-based flux tuning, and the coherent couplings \(g_{1}\), \(g_{2}\), and \(J\) can be implemented through capacitive, inductive, or tunable-coupler-mediated interactions. In particular, tunable superconducting resonators have already been used to observe EPs, with frequency and damping controlled by a SQUID and a quantum-circuit refrigerator, respectively~\cite{Partanen5052019}. Tunable and switchable couplings between superconducting resonators in the MHz range have also been experimentally demonstrated~\cite{Baust5152015}. In addition to local resonator-based implementations, chiral waveguide-QED architectures provide another promising superconducting-circuit route for engineering directional microwave channels and non-Hermitian effective interactions. Recent experiments have demonstrated deterministic remote entanglement and on-demand directional microwave photon emission using chiral quantum interconnects, while related theoretical work has shown that tunable atomic mirrors in waveguide cavity QED can realize non-Hermitian interactions~\cite{Almanakly2025,Nie2023,Kannan2023}. These developments indicate that superconducting waveguide-QED platforms can provide controllable Markovian channels and directional couplings that may be combined with auxiliary structured reservoirs to emulate the non-Markovian higher-dimensional effective Hamiltonians considered in our work.

A crucial ingredient of our proposal is the structured non-Markovian reservoir. In the theoretical model, the two reservoirs coupled to the two ports of mode \(a\) are characterized by the Lorentzian spectral widths \(\lambda_{\ell}\) and the reservoir-induced external decay rates \(\Gamma_{\ell}\), with \(\ell=1,2\). For an experimental superconducting-circuit implementation, these parameters can be engineered through the pseudomode mechanism discussed in Appendix~B. Namely, a Lorentzian reservoir can be realized by coupling the target resonator to an auxiliary lossy microwave resonator. The auxiliary-resonator linewidth and the target-mode--pseudomode coupling are not additional independent parameters in the numerical spectra; rather, they are derived implementation parameters fixed by the target Lorentzian parameters \(\lambda_{\ell}\) and \(\Gamma_{\ell}\). If the auxiliary pseudomode has linewidth \(\kappa_{x,\ell}\) and couples to the target mode with strength \(G_{\ell}\), the Lorentzian parameters satisfy \(\lambda_{\ell}=\kappa_{x,\ell}/2\) and \(\Gamma_{\ell}=4G_{\ell}^{2}/\kappa_{x,\ell}\) in Eq.~(\ref{RB6}). Equivalently, after converting all angular-frequency quantities to ordinary frequencies \(x/2\pi\), the required target-mode--pseudomode coupling is \(G_{\ell}/2\pi=\{[(\lambda_{\ell}/2\pi)(\Gamma_{\ell}/2\pi)]/2\}^{1/2}\).

This implementation is physically transparent: the target resonator exchanges excitations coherently with the auxiliary pseudomode, while the pseudomode itself decays into an approximately Markovian transmission line. The resulting finite pseudomode lifetime produces the memory kernel of the structured reservoir. In the limit \(\kappa_{x,\ell}\gg G_{\ell},g_{1},g_{2},J\), the memory time becomes very short and the reservoir reduces to a Markovian bath; in contrast, when \(\kappa_{x,\ell}\) is comparable to the coherent system rates, the feedback from the pseudomode gives rise to non-Markovian dynamics. Such engineered lossy modes can be realized using readout resonators, Purcell-filtered resonators, or quantum-circuit-refrigerator-type dissipative elements~\cite{Tan151,Sevriuk082601}. Recent superconducting-circuit experiments have also demonstrated non-Markovian quantum EPs by coupling a Josephson-junction-based qubit to a leaky electromagnetic resonator, which acts as a structured reservoir~\cite{HaoLZ2025}.

To connect our theoretical parameters to experimentally realistic values, Table~\ref{tab:experimental_parameters} summarizes representative target parameters for the two main non-Markovian numerical examples: the EP4 case in Fig.~\ref{fig:Fig2}(a) and the EP5 case in Fig.~\ref{fig:Fig2}(b). All rates and frequencies are quoted as \(x/2\pi\). Since the memory kernel generated by a Lorentzian reservoir decays as \(f_{\ell}(t)\propto e^{-\lambda_{\ell}t}\), the bath correlation time is estimated as \(\tau_{\mathrm{B},\ell}=1/\lambda_{\ell}\). For the EP4 case, \(\lambda_{1}/2\pi=\lambda_{2}/2\pi=3.77~\mathrm{MHz}\), giving \(\tau_{\mathrm{B},1}=\tau_{\mathrm{B},2}\simeq 42~\mathrm{ns}\). The corresponding coherent coupling time at the EP is \(1/g_{1\mathrm{EP4}}\simeq 116~\mathrm{ns}\), where \(g_{1\mathrm{EP4}}/2\pi=1.367~\mathrm{MHz}\). For the EP5 case, the two reservoirs have strongly different memory times, \(\tau_{\mathrm{B},1}\simeq 2.0~\mu\mathrm{s}\) and \(\tau_{\mathrm{B},2}\simeq 54~\mathrm{ns}\), while \(1/g_{1\mathrm{EP5}}\simeq 106~\mathrm{ns}\) for \(g_{1\mathrm{EP5}}/2\pi=1.500~\mathrm{MHz}\). Thus, the reservoir memory is not negligible on the system evolution timescale. This is precisely the regime in which the structured reservoir cannot be replaced by a $\delta$-correlated Markovian bath. Separately, the Markovian reference simulations in Fig.~\ref{fig:Fig3}(c) and (d) use very broad Lorentzian reservoirs, \(\lambda_{1}/2\pi=\lambda_{2}/2\pi=200~\mathrm{MHz}\), corresponding to \(\tau_{\mathrm{B}}\simeq 0.8~\mathrm{ns}\). This bath correlation time is two orders of magnitude shorter than the coherent coupling time. This quantitative separation is consistent with the reduction of the higher-order non-Markovian EPs to lower-order EPs in the Markovian reference cases.

\begin{table*}[!t]
\caption{Representative superconducting-circuit parameters for observing the non-Markovian-induced higher-order EPs. All rates and frequencies are quoted as \(x/2\pi\). The EP4 column corresponds to the non-Markovian spectrum in Fig.~\ref{fig:Fig2}(a), and the EP5 column corresponds to the non-Markovian spectrum in Fig.~\ref{fig:Fig2}(b), further analyzed in Fig.~\ref{fig:Fig14}(a) and (b). The first rows list the principal parameters used in the non-Markovian numerical examples, evaluated at the corresponding EPs where applicable. The side-mode detunings are listed explicitly; the central-mode detuning in the EP5 case and the reservoir-induced decay rates \(\Gamma_{\ell}\) are determined by the pseudo-Hermiticity constraints. The pseudomode linewidths are implementation parameters obtained from the pseudomode mapping of the Lorentzian reservoirs, with \(\kappa_{x,\ell}=2\lambda_{\ell}\). Once the reservoir-induced decay rates \(\Gamma_{\ell}\) are fixed, the corresponding auxiliary couplings follow from \(G_{\ell}=\sqrt{\lambda_{\ell}\Gamma_{\ell}/2}\); they are not additional independent parameters used in the numerical spectra. The bath correlation time is estimated as \(\tau_{\mathrm{B},\ell}=1/\lambda_{\ell}\), consistent with the Lorentzian memory kernel \(f_{\ell}(t)\propto e^{-\lambda_{\ell}t}\). The loaded pseudomode quality factor is estimated at a representative microwave frequency of \(5~\mathrm{GHz}\). The Markovian comparison panels in Figs.~3(c), 3(d), 4(c), and 4(d) use the separate parameter sets stated in their captions and are not included as implementation columns here.}
\label{tab:experimental_parameters}
\begin{ruledtabular}
\begin{tabular}{lcc}
Quantity & EP4 in Fig.~2(a) & EP5 in Fig.~2(b) \\
\hline
Target-mode frequency\textsuperscript{a} & \(4{-}8~\mathrm{GHz}\) & \(4{-}8~\mathrm{GHz}\) \\
Critical coupling \(g_{1\mathrm{EP4}}/2\pi\)\textsuperscript{b} & \(1.367~\mathrm{MHz}\) & \(1.500~\mathrm{MHz}\) \\
Coupling \(g_{2}/2\pi\)\textsuperscript{b} & \(g_{2}=g_{1\mathrm{EP4}}=1.367~\mathrm{MHz}\) & \(0.589~\mathrm{MHz}\) \\
Inter-mode coupling \(J/2\pi\)\textsuperscript{b} & \(0\) & \(0.0058~\mathrm{MHz}\) \\
Side-mode decay \(\gamma_{1}/2\pi\)\textsuperscript{c} & \(1.885~\mathrm{MHz}\) & \(2.000~\mathrm{MHz}\) \\
Side-mode decay \(\gamma_{2}/2\pi\)\textsuperscript{c} & \(1.885~\mathrm{MHz}\) & \(1.000~\mathrm{MHz}\) \\
Side-mode detuning \(\Delta_{b_1}/2\pi\) & \(0.866~\mathrm{MHz}\) & \(0.3935~\mathrm{MHz}\) \\
Side-mode detuning \(\Delta_{b_2}/2\pi\) & \(-0.866~\mathrm{MHz}\) & \(-0.787~\mathrm{MHz}\) \\
Central-mode detuning \(\Delta_{a}/2\pi\) & \(0\) & Eq. (\ref{constraint}) \\
Reservoir width \(\lambda_{1}/2\pi\)\textsuperscript{d} & \(3.77~\mathrm{MHz}\) & \(0.079~\mathrm{MHz}\) \\
Reservoir width \(\lambda_{2}/2\pi\)\textsuperscript{d} & \(3.77~\mathrm{MHz}\) & \(2.921~\mathrm{MHz}\) \\
Memory time \(\tau_{\mathrm{B},1}\)\textsuperscript{d} & \(42~\mathrm{ns}\) & \(2.0~\mu\mathrm{s}\) \\
Memory time \(\tau_{\mathrm{B},2}\)\textsuperscript{d} & \(42~\mathrm{ns}\) & \(54~\mathrm{ns}\) \\
Pseudomode linewidth \(\kappa_{x,1}/2\pi\)\textsuperscript{e} & \(7.54~\mathrm{MHz}\) & \(0.158~\mathrm{MHz}\) \\
Pseudomode linewidth \(\kappa_{x,2}/2\pi\)\textsuperscript{e} & \(7.54~\mathrm{MHz}\) & \(5.842~\mathrm{MHz}\) \\
Loaded \(Q_{x,1}\) at \(5~\mathrm{GHz}\)\textsuperscript{f} & \(6.6\times 10^{2}\) & \(3.2\times 10^{4}\) \\
Loaded \(Q_{x,2}\) at \(5~\mathrm{GHz}\)\textsuperscript{f} & \(6.6\times 10^{2}\) & \(8.6\times 10^{2}\) \\
\end{tabular}
\end{ruledtabular}

\vspace{0.5ex}
\begin{minipage}{\textwidth}
\footnotesize
\justifying
\setlength{\parindent}{0pt}
\emergencystretch=2em

\vspace{0.5pt}
\textsuperscript{a} Representative microwave-resonator frequencies, MHz-scale detunings, and flux-tunable superconducting-circuit parameters are standard in circuit-QED platforms~\cite{Xiang6232013,Gu12017,Blais025005,Krantz021318,Partanen5052019}.
\textsuperscript{b} MHz-scale coherent couplings and tunable/switchable resonator-resonator couplings have been demonstrated in superconducting circuits~\cite{Gu12017,Blais025005,Krantz021318,Baust5152015}.
\textsuperscript{c} Engineered damping rates and external linewidths in the MHz range can be realized using tunable losses, overcoupled resonators, Purcell-filtered modes, or quantum-circuit-refrigerator-type elements~\cite{Partanen5052019,Tan151,Sevriuk082601}.
\textsuperscript{d} Lorentzian structured reservoirs and their memory kernels are described by the pseudomode framework and related non-Markovian open-system mappings~\cite{Garraway5522901997,Garraway5546361997,Mazzola800121042009,Tamascelli120030402,Lin18362}; superconducting-circuit realizations of non-Markovian EPs with a leaky resonator acting as a structured reservoir have also been reported~\cite{HaoLZ2025}.
\textsuperscript{e} The pseudomode linewidths are derived from the Lorentzian mapping, \(\kappa_{x,\ell}=2\lambda_{\ell}\). The auxiliary couplings, when needed, are obtained from \(G_{\ell}=\sqrt{\lambda_{\ell}\Gamma_{\ell}/2}\) after fixing the reservoir-induced decay rates \(\Gamma_{\ell}\), and their implementation relies on lossy auxiliary resonators and coherent resonator couplings~\cite{Baust5152015,Tan151,Sevriuk082601,Garraway5522901997,Garraway5546361997,Mazzola800121042009,Tamascelli120030402,Lin18362,HaoLZ2025}.
\textsuperscript{f} The loaded quality factors listed here correspond to deliberately engineered external linewidths and remain within the range accessible in superconducting microwave-resonator platforms~\cite{Gu12017,Blais025005,Krantz021318,Tan151,Sevriuk082601}.

\end{minipage}
\end{table*}

The numbers in Table~\ref{tab:experimental_parameters} are compatible with current superconducting-circuit technology. For example, the EP4 implementation requires deliberately engineered external linewidths of order \(1{-}10~\mathrm{MHz}\), corresponding to loaded quality factors \(Q\sim 10^{3}\) at \(5~\mathrm{GHz}\). Such linewidths are readily obtained by overcoupling resonators to transmission lines, by Purcell-filter engineering, or by using quantum-circuit refrigerators~\cite{Tan151,Sevriuk082601}. The EP5 example additionally requires one relatively narrow pseudomode with \(\kappa_{x,1}/2\pi=0.158~\mathrm{MHz}\), corresponding to \(Q_{x,1}\simeq 3.2\times 10^{4}\) at \(5~\mathrm{GHz}\), which remains well within the range of superconducting microwave resonators~\cite{Gu12017,Blais025005,Krantz021318}. Therefore, the required non-Markovian reservoirs do not require exotic material parameters; they require controlled external coupling and calibration of auxiliary resonator linewidths. The more general examples beyond the pseudo-Hermitian constraint in Fig.~5 also use MHz-scale detunings, couplings, reservoir widths, and engineered linewidths. Their implementation requirements are therefore of the same superconducting-circuit order of magnitude as those summarized in Table~\ref{tab:experimental_parameters}. The CPA condition can be implemented using two phase-coherent microwave tones injected into the two ports coupled to mode \(a\). According to Eq.~(\ref{eq:CPA condition-1}), the two input fields must have a fixed relative phase and an amplitude ratio determined by \(a_{1}^{(\mathrm{in})}(\omega_{\mathrm{CPA}})/a_{2}^{(\mathrm{in})}(\omega_{\mathrm{CPA}})=\kappa_{1}(-\omega_{\mathrm{CPA}})/\kappa_{2}(-\omega_{\mathrm{CPA}})\). In practice, this can be achieved with a single microwave source split into two arms, followed by calibrated attenuators and phase shifters, or equivalently by IQ mixers (i.e., In-Phase and Quadrature mixers), to set the required relative amplitude and phase. Importantly, the effective gain in our CPA description does not require a physical gain medium with population inversion; it is induced by coherent interference of the input fields in the scattering response. This avoids additional gain noise and makes the proposal closer to standard microwave scattering experiments~\cite{Sun9032014,Zhang3682017,Wang2612021,Zhang2022020}. The total output spectrum \(|T_{\mathrm{tot}}(\omega)|^{2}\) can be measured using heterodyne detection. The EP can then be identified by tracking the CPA-induced minima in the output spectrum while tuning 
\(g_1\), \(g_2\), the detunings, or the engineered linewidths, and by fitting the measured spectral response to the effective Hamiltonian.

We now discuss the main experimental challenges and their mitigation. First, intrinsic resonator losses should be smaller than the engineered linewidths listed in Table~\ref{tab:experimental_parameters}, and residual thermal photon populations should be sufficiently suppressed. This requirement on the loss rates is realistic because high-quality superconducting resonators can have intrinsic linewidths well below the deliberately engineered MHz-scale external losses. For a \(5~\mathrm{GHz}\) resonator with internal quality factor \(Q_{i}=10^{5}{-}10^{6}\), the intrinsic linewidth is only \(\kappa_{i}/2\pi\simeq 5{-}50~\mathrm{kHz}\), which is much smaller than the listed side-mode decay rates \(\gamma_{1,2}\) and pseudomode linewidths \(\kappa_{x,\ell}\) used in our proposal. The reservoir-induced rates \(\Gamma_{\ell}\), determined by the pseudo-Hermiticity constraints, can be calibrated independently through the target-mode-pseudomode couplings~\cite{Gu12017,Blais025005,Krantz021318}. Thermal photons can be suppressed by operating at dilution-refrigerator temperatures and by using standard microwave attenuation, filtering, and infrared shielding. Second, flux noise in SQUID-based tunable elements can induce slow fluctuations of detunings and couplings. This can be mitigated by operating near flux-insensitive points when possible, using moderate tuning ranges rather than extreme bias points, and performing in situ calibration of the effective Hamiltonian before the EP scan~\cite{Partanen5052019,Baust5152015}. Third, fabrication tolerances in resonator frequencies and coupling capacitances do not need to place the device directly at the EP. The relevant conditions are imposed after fabrication by flux tuning, tunable couplers, and engineered external decay channels. Moderate static parameter offsets can therefore be absorbed into the calibrated detunings and coupling strengths. Fourth, CPA requires stable relative phase and amplitude between the two input tones. Since both tones can be derived from the same microwave source, residual phase drift is mainly an interferometric calibration issue and can be corrected by measuring the off-resonant scattering background. Finally, for EP5, EP6, and EP7, additional distinguishable non-Markovian reservoirs require additional auxiliary pseudomodes. The main challenge is unwanted cross-talk between these auxiliary modes, which can be reduced by frequency separation, dedicated feedlines, Purcell-filter design, and independent calibration of each linewidth~\cite{Tan151,Sevriuk082601,HaoLZ2025}. These considerations suggest that the EP4 configuration is the most direct near-term implementation, while the EP5--EP7 cases are accessible by systematically adding independently engineered Lorentzian reservoirs.

\section{Conclusion and discussion}\label{Discussion and conclusions}

To summarize, we have studied a pseudo-Hermitian optical system consisting of three coupled cavities interacting with non-Markovian structured reservoirs. We have demonstrated that memory effects rooted in non-Markovianity effectively expand the dimensionality of the effective Hamiltonian, thereby enabling the emergence of higher-order EPs. We have also observed that the pseudo-Hermitian system, with an effective gain induced by CPA, enables higher-order EPs to be directly identified in the output spectrum. We have systematically classified higher-order EPs under different parameter-space symmetries and structured-reservoir configurations, and have revealed a general rule consistent with pseudomode theory:~coupling to multiple distinguishable non-Markovian structured reservoirs with Lorentzian spectra enlarges the Hamiltonian dimension, which in turn increases the parametric complexity and yields higher-order EPs. We have further shown that these non-Markovianity-induced higher-order EPs admit a topological characterization as defects in the relevant pseudo-Hermitian parameter space, with quantized charges described by resultant winding numbers. Moreover, we have pointed out that the non-Markovian mechanism behind higher-order EPs extends beyond pseudo-Hermitian systems to generic non-Hermitian quantum systems. In addition, possible experimental schemes based on superconducting circuits have been discussed, providing feasible routes for observing our findings.

The results of this work are nevertheless restricted to the rotating-wave approximation and to non-Markovian structured reservoirs whose spectral densities are described by a Lorentzian shape. In the future, it is important to analyze scenarios beyond the rotating-wave approximation~\cite{Shen7072022,Shen042121,Shen0338052017,Shen013826}, incorporating counter-rotating terms such as anisotropic non-rotating-wave interactions~\cite{Xie0210462014,Chen0437082021,Ai042116,Lu0543022007}, which will broaden the applicability and generality of our theory. Another interesting direction is to realize arbitrary-order EPs within a single environment by engineering non-Markovian structured reservoirs with a non-Lorentzian spectral density. Specifically, our work demonstrates that each additional distinguishable Lorentzian structured reservoir contributes one auxiliary pseudomode dimension to the effective Hamiltonian, thereby increasing the accessible EP order under appropriate parameter constraints. However, by suitably engineering a single non-Markovian reservoir with a non-Lorentzian spectral density, it may be possible to realize arbitrary-order EPs. An extreme example is that a single non-Markovian structured reservoir with an Ohmic spectral density can give rise to EPs of arbitrarily high order, up to infinity. The analysis of higher-order EPs induced by non-Markovian structured reservoirs offers new insights and effective pathways for designing quantum devices with enhanced sensitivity under realistic conditions.

\section*{Author contributions}
This project was conceived by H. Z. Shen, Zhi-Guang Lu, Yan-Hui Zhou, and C. Shang. The methodology was developed by H. Z. Shen, X. C. Zhang, L. Y. Ning, Yan-Hui Zhou, and C. Shang. The project was supervised by H. Z. Shen, Yan-Hui Zhou, and C. Shang. The original draft of the manuscript was written by H. Z. Shen, X. C. Zhang, and C. Shang. The manuscript was reviewed and edited by H. Z. Shen, X. C. Zhang, L. Y. Ning, Zhi-Guang Lu, Yan-Hui Zhou, and C. Shang.

\begin{acknowledgments}
We thank the referees for their constructive comments that helped
improve our work. C. S. thanks Dr.~Aziza Almanakly from MIT for valuable discussions on the experimental feasibility, both in person during the RIKEN visit and subsequently by email. C. S. also thanks Dr.~Hui Wang from Franco Nori's group for helpful discussions. This work was supported by the Science and Technology Development Plan Project of Jilin Province (Grant No.~20250102007JC), and the National Natural Science Foundation of China under Grant No.~12274064 and No.~12374333. C. S. acknowledges financial support from the China Scholarship Council, the Japanese Government (Monbukagakusho-MEXT) Scholarship (Grant No.~211501), the RIKEN Junior Research Associate Program, and the Hakubi Projects of RIKEN.
\end{acknowledgments}

\section*{Conflicts of interest}
The authors declare no conflicts of interest.

\section*{Data availability statement}
The data that support the findings of this study are available from the corresponding author upon reasonable request.

\section*{Code availability statement}
Code used to generate data in this study is available from the corresponding author upon reasonable request.\\

\noindent \textbf{Correspondence} and requests for materials should be addressed to C. Shang or H. Z. Shen.

\appendix
\section{Dynamics and the non-Markovian input-output relation.}\label{non-Markovian input–output}

For Hamiltonian~(\ref{eq:H_T}), the corresponding Heisenberg equations are given by
\begin{equation}
\begin{aligned}
\dot a =  &- i{\Delta _a}a - i{g_1}{b_1} - i{g_2}{b_2}\\ 
&- \sum_k {C_k^*} {c_k} - \sum_s {D_s^*} {d_s},\\
{{\dot b}_1} =  &- i({\Delta _{{b_1}}} - i{\gamma _1}){b_1} - i{g_1}a - iJ{b_2},\\
{{\dot b}_2} =  &- i({\Delta _{{b_2}}} - i{\gamma _2}){b_2} - i{g_2}a - iJ{b_1},\\
{{\dot c}_k} =  &- i{\Omega _k}{c_k} + {C_k}a, \ {{\dot d}_s} =  - i{\varpi _s}{d_s} + {D_s}a. \label{eq:dynamics}
\end{aligned}
\end{equation}
Solving Eq.~(\ref{eq:dynamics}) yields the time-dependent environment operators for $t \ge 0$ in the form
\begin{equation}
\begin{aligned}
{c_k}(t) = {c_k}(0){e^{ - i{\Omega _k}t}} + {C_k}\int_0^t {d\tau a} (\tau ){e^{ - i{\Omega _k}(t - \tau )}},\\
{d_s}(t) = {d_s}(0){e^{ - i{\varpi _s}t}} + {D_s}\int_0^t {d\tau a} (\tau ){e^{ - i{\varpi _s}(t - \tau )}},\label{eq:environmental-operator}
\end{aligned}
\end{equation}
which can be divided into two components. The first term in ${c_k}(t)$ or ${d_s}(t)$ corresponds to the evolution of the non-Markovian environmental field, while the second term reflects the feedback of non-Markovian effects from the environment to cavity $a$. Further, substituting Eq.~(\ref{eq:environmental-operator}) into Eq.~(\ref{eq:dynamics}) yields the integro-differential equation for the cavity operator
\begin{equation}
\begin{aligned}
\dot a =  &- i{\Delta _a}a - i{g_1}{b_1} - i{g_2}{b_2} + {K_1}(t) + {K_2}(t)\\
 &- \int_0^t {d\tau a} (\tau )\left[ {{f_1}(t - \tau ) + {f_2}(t - \tau )} \right],\label{eq:cavity-operator}
\end{aligned}
\end{equation}
where ${K_1}(t) =  - \sum\nolimits_k {C_k^*{c_k}(0)\exp ( { - i{\Omega _k}t} )} $ $= \int_{ - \infty }^\infty  {d\tau \kappa _1^*} (t - \tau )a_1^{({\rm{in}})}(\tau )$ and ${K_2}(t) =  - \sum\nolimits_s {D_s^*} {d_s}( 0 ){e^{ - i{\varpi _s}t}}$ $ = \int_{ - \infty }^\infty  {d\tau \kappa _2^*} (t - \tau )a_2^{({\rm{in}})}(\tau )$.~Herein, the input field operators are defined as $a_1^{(\rm{in})}\left( t \right) =  - {1}/{{\sqrt {2\pi } }}\left(\sum\nolimits_k {{c_k}} (0){e^{ - i{\Omega _k}t}}\right),~a_2^{(\rm{in})}\left( t \right) =  -{1}/{{\sqrt {2\pi } }}\left(\sum\nolimits_s {{d_s}} (0){e^{ - i{\varpi _s}t}}\right),$
and the impulse response functions in the continuum limit are denoted by
\begin{equation}
\begin{aligned}
{\kappa _1}(t - \tau ) =& \frac{1}{{\sqrt {2\pi } } }\int_{ - \infty }^{ + \infty }d\omega {{e^{i\omega (t - \tau )}}} C(\omega ), \\
{\kappa _2}(t - \tau ) =& \frac{1}{{\sqrt {2\pi } }}\int_{ - \infty }^{ + \infty }d\omega' {{e^{i\omega' (t - \tau )}}} D(\omega' ),\label{eq:continuum limit}
\end{aligned}
\end{equation}
where ${C_k} \to C\left( \omega  \right)$ and ${D_s} \to D\left( {\omega '} \right)$. Notably, the correlation functions $f_1(t)$ and $f_2(t)$ in Eq.~(\ref{eq:cavity-operator}) play a central role by acting as memory kernels, which are given by
\begin{equation}
\begin{aligned}
{f_1}(t) &= \int_{ - \infty }^{ + \infty } {{S_1}(\omega ){e^{ - i\omega t}}d\omega,} \\ {f_2}(t) &= \int_{ - \infty }^{ + \infty } {{S_2}(\omega' ){e^{ - i\omega' t}}d\omega',}\label{eq:memory kernel}
\end{aligned}
\end{equation}
where $S_1(\omega)=|C(\omega)|^{2}$ and $S_2(\omega')=|D(\omega')|^{2}$ are referred to as the spectral densities of the reservoirs. Similarly, for $0 \le t \le t_1$, the solution for the output field operator reads
\begin{equation}
\begin{aligned}
{a}^{(\rm{out})}_{1}(t)&=\frac{1}{\sqrt{2\pi}}\sum_k{c}_{k}(t_{1})e^{-i\Omega_{k}(t-t_{1})},\\
{a}^{(\rm{out})}_{2}(t)&=\frac{1}{\sqrt{2\pi}}\sum_s{d}_{s}(t_{1})e^{-i{{\varpi _s}}(t-t_{1})}.\label{eq:output-operator}
\end{aligned}
\end{equation}
Combining Eq.~(\ref{eq:cavity-operator}) to Eq.~(\ref{eq:output-operator}) and taking the limit $t_1 \to t$, we obtain the non-Markovian input–output relation for the port of cavity $a$ as follows:
\begin{align}
{a}_{1(2)}^{(\rm{in})}(t)+{a}_{1(2)}^{(\rm{out})}(t)=\int_{0}^{t}\kappa_{1{(2)}}(\tau-t){a}(\tau)d\tau.\label{eq:input-output relation}
\end{align}

Taking the impulse response functions given in Eq.~(\ref{eq:continuum limit}) as $\kappa_1(t)=\lambda_1\sqrt{\Gamma_{1}}e^{\lambda_1 t}\vartheta (-t)$ and $\kappa_2(t)=\lambda_2\sqrt{\Gamma_{2}}e^{\lambda_2 t}\vartheta(-t)$, we obtain the correlation functions $f_1(t)=\lambda_1\Gamma_{1}e^{-\lambda_1|t|} /2$ and $f_2(t)=\lambda_2\Gamma_{2}e^{-\lambda_2|t|} /2$ from Eq.~(\ref{eq:memory kernel}), where $\vartheta \left( { - t} \right)$ is the step function, defined as $\vartheta \left( { - t} \right)=1$ for $t \le 0$ and $\vartheta \left( { - t} \right) = 0$ for $t > 0$. After Fourier transform to Eq.~(\ref{eq:continuum limit}), the corresponding spectral response functions read $C(\omega)=\sqrt{{\Gamma_{1}}/{2\pi}}[{\lambda_{1}}/({\lambda_{1}-i\omega})]$~and~$\ D(\omega')=\sqrt{{\Gamma_{2}}/{2\pi}}{\lambda_{2}}/[({\lambda_{2}-i\omega'})]$, 
where $\lambda_{1}$ and $\lambda_{2}$ represent the spectral widths of two non-Markovian environments, while $\Gamma_{1}$ and $\Gamma_{2}$ denote the cavity dissipation rates coupled to the two ports. Consequently, the Lorentzian spectral densities~\cite{Shen0121562017} take the form
\begin{align}
S_{1}(\omega)=\frac{\Gamma_{1}}{2\pi}\frac{\lambda_{1}^{2}}{\lambda_{1}^{2}+\omega^{2}}, \ S_{2}(\omega')=\frac{\Gamma_{2}}{2\pi}\frac{\lambda_{2}^{2}}{\lambda_{2}^{2}+(\omega')^{2}},
\label{eq:Lorentzian}
\end{align}
which characterize a Gaussian Ornstein--Uhlenbeck process \cite{Uhlenbeck8231930}. Intuitively, the memory effect associated with non-Markovian reservoirs disappears as the spectral widths $\lambda_{1}$ and $\lambda_{2}$ approach infinity, leading to a return to the Markovian case~\cite{Walls1994,Scully1997}. Notably, the above derivation for the impulse response and correlation functions can also be carried out using the pseudomode theory with details provided in Appendix~\ref{Pseudomode analysis}.

\section{Pseudomode analysis}\label{Pseudomode analysis}
Readers primarily interested in the effective Hamiltonian may skip this appendix, as the main results are identical to those in Appendix~\ref{non-Markovian input–output}. Within the pseudomode framework, a structured bath is replaced by a set of locally damped auxiliary modes that reproduce its spectral density \cite{Alford250916377}. As an example, for a cavity mode, this Markovian embedding realizes a controllable Lorentzian bath spectral density in a non-Markovian environment \cite{Jack0438032001,Barnett1997,Garraway5522901997,Garraway5546361997,Man900621042014,Mazzola800121042009,Pleasance960621052017,Tamascelli120030402}.
We introduce a system composed of a cavity mode (eigenfrequency ${\omega _n}$) coupled to a pseudomode (eigenfrequency $\omega _x$). The Hamiltonian for this system in a rotating frame with $\omega_x$ is
\begin{eqnarray}
{{H}_{S}} =  {\Delta_n}{{ n}^\dag } n  + {G}( n{{ x}^\dag } + {{ n}^\dag } x),
\label{xuanzhuanqianHS}
\end{eqnarray}
where the first term on the right-hand side is the free Hamiltonians of the cavity. The annihilation operators $ n$ and $ x$ satisfy the bosonic commutation relations $[ n,  {n}^\dag ] = [ x,  {x}^\dag ] = 1$. $\Delta_n=\omega_n-\omega_x$ denotes the detuning between cavity and pseudomode.
The second term in Eq.~(\ref{xuanzhuanqianHS}) corresponds to the tunneling coupling between the cavity mode and the pseudomode with the coupling strength \(G\).
The corresponding total Hamiltonian including the Markovian environment reads
\begin{equation}
\begin{aligned}
{{ H_T}} = {{ H}_S} + {{ H}_R} + {{ H}_I},
\label{HIHS}
\end{aligned}
\end{equation}where ${{ H}_I}  = i\sum\nolimits_k {{V_k}} ( q_k^\dag  x - {{ x}^\dag }{{ q}_k})$ denotes the interaction Hamiltonian between the pseudomode and the Markovian environment, with ${V_k} = \sqrt {\kappa_{x} /2\pi }$ and $\kappa_{x}$ representing the coupling strength and decay rate, respectively. The free Hamiltonian of the environment is given by ${ H_R}   = \sum\nolimits_k {{(\omega _k-\omega _x)} q_k^\dag { q_k}}$, which satisfies the commutation relation $[{ q_k},  q_{k'}^\dag ] = {\delta _{kk'}}$. According to Eq.~(\ref{HIHS}), the Heisenberg-Langevin equations under the Markovian approximation take the form \cite{GardinerBerlin2000,Walls1994}
\begin{eqnarray}
\frac{d}{{dt}} n &=&  - i\Delta _n n - i{G} x, \label{dota}\\
\frac{d}{{dt}} x &=&  - i{G} n - \frac{k_x }{2} x - \sqrt {k_x}  {{ q}_{\rm in}}(t).
\label{dotb}
\end{eqnarray}
Solving Eq.~(\ref{dotb}) for $ x\left( t \right)$ gives $  x(t) =  x(0){e^{ - \frac{{k_x} }{2}t}} - i{G}\int_0^t { n(\tau ){e^{ -  \frac{{k_x} }{2}(t - \tau )}}d\tau } - \sqrt {k_x}  \int_0^t {{{ q}_{\rm in}}(\tau ){e^{ - \frac{{k_x} }{2}(t - \tau )}}d\tau }$ with ${ q_{{\rm{in}}}}(t){\rm{ = }}\sum\nolimits_k {{e^{ - i{(\omega _k-\omega _x)}t}}} { q_k}/\sqrt {2\pi }$. Substituting $ x(t)$ into Eq.~(\ref{dota}), we obtain
\begin{equation}
\begin{aligned}
\frac{d}{{dt}} n =  - i\Delta _n n - \int_0^t {\beta (t - \tau ) n(\tau )d\tau }-{R}(t),
\label{Heisenbergequationsofmotion3}
\end{aligned}
\end{equation}
where $ R(t) = i{G} x(0){e^{ -\frac{{k_x} }{2}t}}-i {G}\sqrt {k_x}  \int_0^t {{{ q}_{\rm in}}(\tau ){e^{ - \frac{{k_x} }{2}(t - \tau )}}d\tau }$ is the operator for the non-Markovian composite environment, which comprises the pseudomode and its Markovian environment, and $\beta(t) = G^2{e^{ -  \frac{k_x }{2}t}}$ is the corresponding correlation function.
The Lorentzian spectrum density $S(\omega ) = \Gamma {\lambda ^2}/[2\pi ({\lambda ^2} + {\omega ^2})]$ corresponding to the correlation function $\beta(t) = G^2{e^{ -  \frac{{k_{x}} }{2}t}} \equiv \int {S(\omega )} {e^{ - i\omega t}}d\omega $ in Eq.~(\ref{Heisenbergequationsofmotion3}) results in
\begin{equation}
\begin{aligned}
\lambda = \frac{k_{x}}{2},~\Gamma  = \frac{4G^2}{k_{x}},
\label{RB06}
\end{aligned} 
\end{equation}
and
\begin{align}
\beta(t-\tau) = \frac{1}{2}\Gamma\lambda e^{-\lambda(t-\tau)}.
\label{RB07}
\end{align}
Defining the expectation values of the operators as $n \coloneqq \left\langle{n}\right\rangle$, $x \coloneqq \langle {x} \rangle$, $q_k \coloneqq \langle {q}_k \rangle$, $q_{\rm in} \coloneqq \langle {q}_{\rm in} \rangle$, $R(t) \coloneqq \langle {R}(t) \rangle$, $a_{\rm in} \coloneqq \langle {a}_{\rm in} \rangle$, $K(t) \coloneqq \langle {K}(t) \rangle$, $c \coloneqq \left\langle{c}\right\rangle$, and considering
\begin{equation}
\begin{aligned}
K(t) = -R(t),
\end{aligned}
\end{equation}
we obtain
\begin{equation}
\begin{aligned}
q_{\rm in}(t)&=\frac{k_x}{2}c(t)+c^{\prime}(t),
\label{R1}
\end{aligned}
\end{equation}
\begin{equation}
\begin{aligned}
x=\frac{i}{G}\int_{-\infty}^{+\infty}h^{*}(-\tau)a_{\rm in}(\tau)d\tau,
\label{R2}
\end{aligned}
\end{equation}
with
\begin{equation}
\begin{aligned}
c(t) = \frac{{{G}x{e^{ - \frac{k_x }{2}t}} - iK(t)}}{{{G}\sqrt {k_x}  }},
\label{R3}
\end{aligned}
\end{equation}
\begin{equation}
\begin{aligned}
K(t)=\int_{-\infty}^{+\infty}h^{*}(t-\tau)a_{\rm in}(\tau)d\tau.
\label{R4}
\end{aligned}
\end{equation}
The value of $x=\mathrm{Tr}[{x}(0)\rho_x(0)]$ is determined by the initial state $\rho_x(0)$ of the pseudomode. If $x$ takes
\begin{equation}
\begin{aligned}
x=i\frac{\sqrt{\Gamma}\lambda}{G}\int_{0}^{+\infty}e^{-\lambda\tau}a_{\rm in}(\tau)d\tau,
\label{R5}
\end{aligned}
\end{equation}
then by comparing Eq.~(\ref{R2}) and Eq.~(\ref{R5}), we obtain
\begin{align}
\!\! h(t)\!=\!\sqrt{\Gamma}\lambda e^{\lambda t}\vartheta(-t).
\label{R06}
\end{align}
Considering the interaction of the cavity with two dissipative pseudomodes (corresponding to two non-Markovian Lorentzian environments) based on the above analysis, the Lorentzian spectral densities $S_1(\omega)$ and $S_2(\omega)$ in the correlation functions $\beta_1(t) = G_1^2{e^{ -  \frac{{k_{x,1}} }{2}t}} \equiv \int {S_1(\omega )} {e^{ - i\omega t}}d\omega $ and $\beta_2(t) = G_2^2{e^{ -  \frac{{k_{x,2}} }{2}t}} \equiv \int {S_2(\omega )} {e^{ - i\omega t}}d\omega $ in Eq.~(\ref{Heisenbergequationsofmotion3}) impose those in Eq.~(\ref{eq:Lorentzian}), where
\begin{equation}
\begin{aligned}
\lambda_1 = \frac{k_{x,1}}{2},~\lambda_2 = \frac{k_{x,2}}{2},~\Gamma_1  = \frac{4G_1^2}{k_{x,1}},~\Gamma_2  = \frac{4G_2^2}{k_{x,2}},
\label{RB6}
\end{aligned} 
\end{equation}
which leads to
\begin{equation}
\begin{aligned}
\beta_{1}(t-\tau) =& \frac{1}{2}\Gamma_{1}\lambda_{1}e^{-\lambda_{1}(t-\tau)},\\
\beta_{2}(t-\tau) =& \frac{1}{2}\Gamma_{2}\lambda_{2}e^{-\lambda_{2}(t-\tau)},
\label{RBz7}
\end{aligned}
\end{equation}
and
\begin{align}
\!\!\! h_1(t)\!=\!\sqrt{\Gamma_1}\lambda_1 e^{\lambda_1 t}\vartheta(-t),
h_2(t)\!=\!\sqrt{\Gamma_2}\lambda_2 e^{\lambda_2 t}\vartheta(-t),
\label{R6}
\end{align}
corresponding to $\kappa_1(t)$ and $\kappa_2(t)$ below Eq.~(\ref{eq:input-output relation}). Here, ${n}$ represents ${a}$, $b_1$, or $ b_2$. 
Therefore, when Eqs.~(\ref{RB07}), (\ref{R1}), (\ref{R2}), and (\ref{R5}) are satisfied simultaneously, we demonstrate that the equations derived from the Markovian pseudomode method are completely consistent with Eq.~(\ref{eq:cavity-operator}) in the non-Markovian regime.

\section{General case of two different non-Markovian structured reservoirs}\label{General case-double}
Defining $\mu_1{Y}_1(t) = -{K}_1(t)+\int_{0}^{t}{a}(\tau)f_1(t-\tau)d\tau $ and $ \mu_2{Y}_2(t) = -{K}_2(t)+\int_{0}^{t}{a}(\tau)f_2(t-\tau)d\tau$ with $\mu_{1(2)}\equiv\sqrt{{\lambda_{1(2)} \Gamma_{1(2)}}/{2}}$ for the general case ($\lambda_1 \ne \lambda_2$ and $\Gamma_1 \ne \Gamma_2$), and considering $\hat{a}_{\nu}^{(in)}(t)=\int_{0}^{t}\kappa_{\nu}(\tau-t)\hat{a}(\tau)d\tau$ in CPA occurring, Eq.~(\ref{eq:system-dynamics}) becomes
\begin{align}
\frac{d}{dt}{a}(t) =&-i\Delta_{c}{a}(t)-ig_1{b_{1}}(t)-ig_2{b}_{2}(t)
\nonumber\\&-\mu_1{Y}_1(t)-\mu_2{Y}_2(t),\nonumber\\\frac{d}{dt}{b}_{1}(t) =&-i(\Delta_{1}-i\gamma_{1}){b}_{1}(t)-ig_1{a}(t)-iJ{b_{2}}(t),
\nonumber\\\frac{d}{dt}{b}_{2}(t) =&-i(\Delta_{2}-i\gamma_{2}){b}_{2}(t)-ig_2{a}(t)-iJ{b_{1}}(t),
\label{dotab12YNonMarkovian}\\\frac{d}{dt}{Y}_1(t)=&\lambda_1{Y}_1(t)+\mu_1{a}(t),
\nonumber\\\frac{d}{dt}{Y}_2(t)=&\lambda_2{Y}_2(t)+\mu_2{a}(t),
\nonumber
\end{align}
or in the matrix form
\begin{equation}
\dot {\mathbf{N}}_5=-i{H}_{5}\mathbf{N_5},
\end{equation}
where $\mathbf{N_5}=({a},{b}_{1},{b}_{2},{Y_1},{Y_2})^{T}$. The effective non-Hermitian Hamiltonian ${H}_{5}$ reads
\begin{equation}
\begin{aligned}
\! {H}_{5}=\begin{pmatrix}\Delta_{a}&g_1&g_2&-i\mu_1&-i\mu_2\\g_1&\Delta_{b_1}-i\gamma_{1}&J&0&0\\g_2&J&\Delta_{b_2}-i\gamma_{2}&0&0\\i\mu_1&0&0&i\lambda_1&0\\i\mu_2&0&0&0&i\lambda_2\end{pmatrix},
\label{5*5NM}
\end{aligned}
\end{equation}
where $\mu_1=\sqrt{{\lambda_1 \Gamma_1}/{2}}$ and $\mu_2=\sqrt{{\lambda_2 \Gamma_2}/{2}}$.
The characteristic equation for the eigenvalue $x$ can be written as
\begin{equation}
\begin{aligned}
a'x^{5}+b'x^{4}+c'x^{3}+d'x^{2}+e'x+f'=0.
\end{aligned}\label{x5}
\end{equation}
For the quintic equation, we define the discriminants $\Delta_{b_1}\coloneqq B'^{2}-4A'C'$ and $\Delta_{b_2}\coloneqq P'^{2}-4L'^{5}$. Then, Eq.~(\ref{x5}) has a quintuple real root if and only if $B'^{2}-4A'C'=0$ and $P'^{2}-4L'^{5}=0$ hold simultaneously. Here, $A'$, $B'$ and $C'$ are given by $A'=F'^{2}-12E'^{2}L', B'=6F'^{3}-64E'^{2}F'L'-72E'^{3}M', C'=3F'^{4}-24E'^{2}F'^{2}L'-48E'^{3}F'M'-80E'^{4}L'^{2},
$ where $E'=2G'^2L'^2-2G'^2N'+3G'H'M'-4H'^2L'-G'JL'$, $F'=G'^2P'+3G'JM'-4H'JL'$, $G'=4L'^{3}-9M'^{2}+8L'N'$, $H'=10L'^{2}M'-6M'N'+L'P'$, and
 $J=4L'^{4}-4L'^{2}N'+3M'P'$. Additionally, \{$A', B', C'$\} are also related to \{$L'$, $M'$, $N'$, $P'$\} with
\begin{align}
L'&= 2b'^{2}-5a'c', \ M'= 4b'^{3}-15a'b'c'+25a'^{2}d',\label{64}\\
N'&=7b'^{4}+25a'^{2}c'^{2}-35a'b'^{2}c'+50a'^{2}b'd'-125a'^{3}e',\nonumber\\
P'&= 4b'^{5}\!-25a'b'^{3}c'\!+125a'^{2}b'^{2}d'\!-625a'^{3}b'e'\!+3125f'a'^{4}.\nonumber
\end{align}
When $L'=M'=N'=P'=0$, Eq.~(\ref{x5}) has a quintuple real root, where $x_1=x_2=x_3=x_4=x_5$, and 
\begin{align}
x_5 =-\frac {b'}{5a'}=-\frac {c'}{2b'}=-\frac{ d'}{c'}=-\frac{2e'}{d'}=-\frac{5f'}{e'}.\label{x_1_2_3_4_5}
\end{align}
Following this, we will explore the determination of the parametric conditions necessary to guarantee the pseudo-Hermiticity of the effective Hamiltonian ${H}_{5}$ in Eq.~(\ref{5*5NM}). By following the procedure described in Ref.~\cite{Mostafazadeh2052002}, ${H}_{5}$ acquires pseudo-Hermitian properties provided that its eigenvalues satisfy a requirement, that is, all five eigenvalues must be real. 
To verify this condition, we solve the equation $\mathrm{Det}({H}_{\rm{5}}-\Omega I)=0$, that is
\begin{align}
\begin{vmatrix}\Delta_a-\Omega&g_1&g_2&-i\mu_1&-i\mu_2\\g_1&M_1&J&0&0\\g_2&J&M_2&0&0\\i\mu_1&0&0&i\lambda_1-\Omega&0\\i\mu_2&0&0&0&i\lambda_2-\Omega\end{vmatrix}=0,
\label{vmatrix5}
\end{align}
where $M_1=(\Delta_{b_1}-i\gamma_1)-\Omega$ and $M_2=(\Delta_{b_2}-i\gamma_2)-\Omega$.
We derive the five eigenvalues based on the energy-spectrum properties described in the pseudo-Hermitian Hamiltonian formalism by considering the complex conjugate of Eq.~(\ref{vmatrix5}), i.e., $\mathrm{Det}({H}_{{\rm{5}}}^{*}-\Omega I)=0$, which yields
\begin{align}
\begin{vmatrix}\Delta_a-\Omega&g_1&g_2&i\mu_1&i\mu_2\\g_1&M_{1}^{*}&J&0&0\\g_2&J&M_{2}^{*}&0&0\\-i\mu_1&0&0&-i\lambda_1-\Omega&0\\-i\mu_2&0&0&0&-i\lambda_2-\Omega\end{vmatrix}=0.
\label{vmarix5x}
\end{align}
Comparing Eqs.~(\ref{vmatrix5}) and (\ref{vmarix5x}) gives the constraint
\begin{widetext}
\begin{align}
&\lambda_1=\gamma_1+\gamma_2-\lambda_2,\quad \Delta_{b_1}=\frac{-\Delta_{b_2} \gamma_2}{\gamma_1},\quad \zeta=-\gamma_2+\frac{\gamma_1^{2}(\gamma_1+\gamma_2-\lambda_2)^{2}}{J^{2}\gamma_1+(\Delta_{b_2}^{2}+\gamma_1^{2})\gamma_2}, \nonumber\\
&\Delta_a=\frac{-2g_1^{2}\Delta_{b_2}\gamma_1^{2}\gamma_2+2g_1g_2J\gamma_1^{2}(\gamma_1+\gamma_2)+\Delta_{b_2}(2g_2^{2}\gamma_1^{2}\gamma_2+J^{2}\gamma_1(\gamma_1^2-\gamma_2^2)+(\omega_2^{2}+\gamma_1^{2})(\gamma_1-\gamma_2)\gamma_2(\gamma_1+\gamma_2))}{\gamma_1(\gamma_1+\gamma_2)(J^{2}\gamma_1+\Delta_{b_2}^{2}\gamma_2+\gamma_1(\gamma_1-\lambda_2)(\gamma_2-\lambda_2))}, \nonumber\\
&\Gamma_1=\frac{2(g_1^{2}\gamma_1^{2}+g_2^{2}\gamma_1\gamma_2+J^{2}\gamma_1(\gamma_1+\gamma_2)+(\Delta_{b_2}^{2}+\gamma_1^{2})\gamma_2(\gamma_1+\gamma_2))-2\gamma_1(\gamma_1+\gamma_2)^{2}\lambda_2+\gamma_1(2(\gamma_1+\gamma_2)-\Gamma_2)\lambda_2^{2}}{\gamma_1(\gamma_1+\gamma_2-\lambda_2)^{2}}, \label{constraint}\\
&\Gamma_2=\frac{2}{\lambda_1-\lambda_2}(-J^{2}+{\Delta_{b_2}}{\Delta_{b_1}}-\gamma_1\gamma_2-\frac{g_1^{2}\gamma_1(J^{2}\gamma_1+\gamma_2(\Delta_{b_2}^{2}-(\gamma_2-\lambda_2)(2\gamma_1+\gamma_2-\lambda_2)))}{(\gamma_1+\gamma_2)(J^{2}\gamma_1+(\Delta_{b_2}^{2}+\gamma_1^{2})\gamma_2)}+\frac{g_2^2\zeta}{\gamma_1+\gamma_2}-\epsilon+\lambda_1\lambda_2), \nonumber\\
&\epsilon=\frac{\Delta_{b_2}(\gamma_1-\gamma_2)(-2g_1^2\Delta_{b_2}\gamma_1^2\gamma_2+2g_1g_2J\gamma_1^2(\gamma_1+\gamma_2)+\Delta_{b_2}(2g_2^2\gamma_1^2\gamma_2
+(\gamma_1^2-\gamma_2^2)(J^2\gamma_1+(\Delta_{b_2}^2+\gamma_1^2)\gamma_2)))(\lambda_1)^2}
{\gamma_1(\gamma_1+\gamma_2)(J^2\gamma_1+(\Delta_{b_2}^2+\gamma_1^2)\gamma_2)(J^2\gamma_1+\Delta_{b_2}^2\gamma_2+\gamma_1(\gamma_1-\lambda_2)(\gamma_2-\lambda_2))}. \nonumber
\end{align}
\end{widetext}

\section{Derivation and discussion of Eq.~(\ref{modified Laplace transform})}\label{Derivation and discussion of Eq16}
Making the modified Laplace transformation to Eq.~(\ref{eq:system-dynamics}) \cite{Shen0121562017,Uchiyama1282009,Shen1222015}
\begin{align}
\eta(\omega)=\int_{0}^{\infty}e^{i\omega t}\eta(t)dt,
\end{align}
with $\exp(i\omega t)\rightarrow \exp(i\omega t-\epsilon t)$ at  $\epsilon\rightarrow0^{+}$ causing $\eta(\omega)$ converge to a finite value with the definitions $\sqrt{2\pi}{a}(t)=\int_{0}^{+\infty}a(\omega)e^{-i\omega t}d\omega$ and $\sqrt{2\pi}{b}_j(t)=\int_{0}^{+\infty}b_j(\omega)e^{-i\omega t}d\omega$, we obtain

\begin{equation}
\begin{aligned}
-i\omega a(\omega)=&-i\Delta_{c} a(\omega)-ig_1b_1(\omega)-ig_2b_2(\omega)\\&+\tilde{\kappa}_{1}(\omega )[a_{in}^{(1)}(\omega)-a_{in}^{(1)}(i\lambda_1)]-a(\omega)f_1(\omega)
\\&+\tilde{\kappa}_{2}(\omega )[a_{in}^{(2)}(\omega)-a_{in}^{(2)}(i\lambda_2)]-a(\omega)f_2(\omega),\label{Langevin1D2}
\\-i\omega b_1(\omega)=&-i(\Delta_1-i\gamma_{1})b_1(\omega)-ig_1a(\omega)-iJb_2(\omega),
\\-i\omega b_2(\omega)=&-i(\Delta_2-i\gamma_{2})b_2(\omega)-ig_2a(\omega)-iJb_1(\omega).
\end{aligned}
\end{equation}

Our goal is to evaluate the impact of the inhomogeneous terms ${ a}_{\rm in}^{(1)}(i{\lambda _1})$ and ${ a}_{\rm in}^{(2)}(i{\lambda _2})$ in Eq.~(\ref{Langevin1D2}), where $\lambda _1$ and $\lambda _2$ denote the spectral widths of the non-Markovian input environments at ports 1 and 2, respectively. Taking ${ a}_{\rm in}^{(1)}(t)$ as an illustrative example, we set $\phi ({\lambda _1},\omega ) = { a}_{\rm in}^{(1)}(i{\lambda _1})/{ a}_{\rm in}^{(1)}(\omega )$. The input field takes two forms: a damped-oscillation form expressed as ${ a}_{\rm in}^{(1)}(t) = {x}{{\rm e}^{ - z t}}\sin ({y}{t^2})$ with $z>0$ and ${y}>0$, and a Gaussian-profile form given by ${ a}_{\rm in}^{(1)}(t) = {x}{{\rm e}^{ - z t^2}}\cos ({y}{t})$. 
These two forms give concrete expressions for $\phi ({\lambda _1},\omega )$. For the damped-oscillation form: $\phi ({\lambda _1},\omega ) = \{ \cos [\frac{{{{\left( {\lambda  + z } \right)}^2}}}{{4y}}][1 - 2fc(\frac{{\lambda  + z }}{{\sqrt {2\pi y} }})] + [1 - 2fs(\frac{{\lambda  + z }}{{\sqrt {2\pi y} }})]\sin [\frac{{{{\left( {\lambda  + z } \right)}^2}}}{{4y}}]\} /\{ \cos [\frac{{{{\left( {z  - {\rm{i}}\omega } \right)}^2}}}{{4y}}][1 - 2fc(\frac{{z  - {\rm{i}}\omega }}{{\sqrt {2\pi y} }})] + [1 - 2fs(\frac{{z  - {\rm{i}}\omega }}{{\sqrt {2\pi y} }})]\sin [\frac{{{{\left( {z  - {\rm{i}}\omega } \right)}^2}}}{{4y}}]\} $. On the other hand, for the Gaussian-profile form: $\phi ({\lambda _1},\omega ) = {e^{\frac{{{\lambda ^2} + {\omega ^2} + 2y(\omega  - i\lambda )}}{{4z }}}}\{ i + {e^{\frac{{{\rm{i}}y\lambda }}{z}}}[i + {\rm {erfi}}(\frac{{y - {\rm{i}}\lambda }}{{2\sqrt z  }})] - {\rm {erfi}}(\frac{{y + {\rm{i}}\lambda }}{{2\sqrt z  }})\} /\{ i + {e^{\frac{{y\omega }}{z }}}[i + {\rm {erfi}}(\frac{{y - \omega }}{{2\sqrt z  }})] - {\rm {erfi}}(\frac{{y + \omega }}{{2\sqrt z  }})\} $, where $fc(\psi ) = \int_0^\psi  {\cos (\pi {t^2}/2)} dt$, $fs(\psi ) = \int_0^\psi  {\sin (\pi {t^2}/2)} dt$, and ${\rm{erfi}}(\psi ) = - i erf (i\psi ) $.
These inhomogeneous terms are determined by the forms of the input field ${ a}_{\rm in}^{(1)}(t)$. In the Markovian approximation, as ${\lambda _1 }\to\infty$, $\phi ({\lambda _1 },\omega )$ approaches zero.

We find that the inhomogeneous terms are incapable of uncovering the characteristics of the systems being probed. For the damped-oscillation form, when $\lambda = {\omega _{\nu}}$, which falls in the non-Markovian regimes, we can estimate $\left| {\phi ({\lambda _1},\omega )} \right|\approx{6\times10^{ - 6}}$, and when $\lambda = 9{\omega _{\nu}}$, where weak non-Markovian effects are present, $\left| {\phi ({\lambda _1},\omega )} \right|\approx{8\times10^{ - 9}}$, where $z = 0.0001{\omega _{\nu}}$, ${y} = 0.00015{\omega _{\nu}}$, and $\omega = {\omega _{\nu}}$. For the Gaussian-profile form, by choosing the same parameters as in the damped-oscillation case, we find that $\left| {\phi ({\lambda _1},\omega )} \right|\approx0$ when $\lambda = {\omega _{\nu}}$ and $\lambda = 9 {\omega _{\nu}}$. The inhomogeneous term ${ a}_{\rm in}^{(1)}(i{\lambda _1})$ is significantly smaller than ${ a}_{\rm in}^{(1)}(\omega )$ and can be neglected for these parameter values. Similar discussions and conclusions can be drawn for the term ${ a}_{\rm in}^{(2)}(i{\lambda _2})$. Consequently, we can safely neglect the influence of the inhomogeneous terms on the system dynamics.
\section{Analysis of parameter-space asymmetry}\label{Asymmetry analysis}
\begin{figure}[h]
   \centerline{
   \includegraphics[width=8.4cm, height=8.0cm, clip]{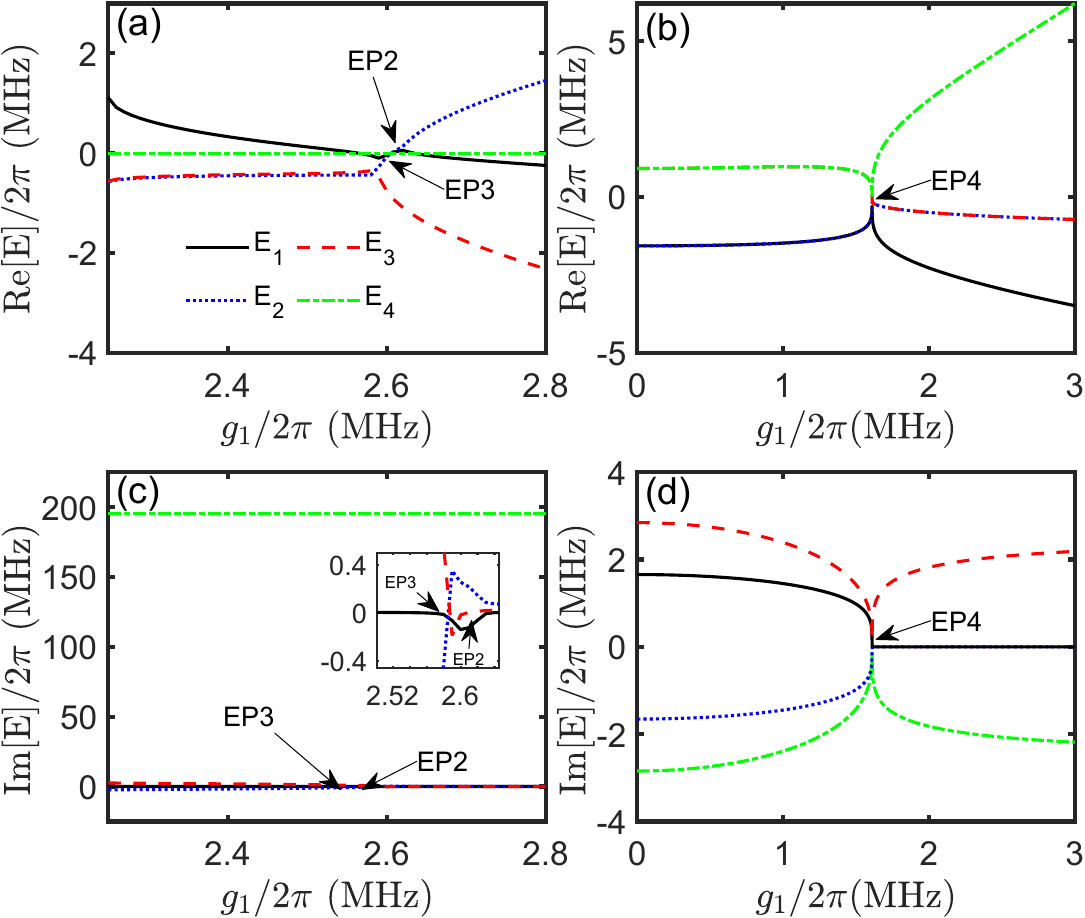}}
   \caption{(a) and (c) respectively plot the evolution of the eigenvalues for the effective Hamiltonian in Eq.~(\ref{eq:effective Hamiltonian}) with Eq. (\ref{eq:pseudo-Hermitian}) as the coupling strength $g_1$ varies at $\gamma_1/2\pi=2\gamma_2/2\pi=3$MHz. The environmental spectral widths are set to $\lambda_1/2\pi = \lambda_2 /2\pi= 200$MHz (Markovian approximation). The parameters chosen are $\Gamma_1/2\pi=\Gamma_2/2\pi=4.5$MHz, $g_2=g_1$, $\Delta_a=0$, $J=0.333g_1$, $\Delta_{b_1}=(0.445g_1^2-\gamma_2^2)$, and $\Delta_{b_2}=-2\Delta_{b_1}$. (b) and (d) present the real and imaginary parts of these eigenvalues with asymmetry parameters $\gamma_1/2\pi=2\gamma_2/2\pi=3$MHz. The remaining parameters are set to $\lambda_1/2\pi = \lambda_2 /2\pi= 4.5$ MHz, $J/2\pi=0.88$MHz, $\Delta_{b_1}/2\pi=0.657$MHz, $\Delta_{b_2}=-2\Delta_{b_1}$, $\Delta_a = {[2g_1^2 J - \Delta_{b_1}(2\gamma_2^2 + J^2 + 2\Delta_{b_1}^2)]}/{(2\gamma_2^2 + J^2 + 2\Delta_{b_1}^2)}$, $\Gamma_1 =\Gamma_2 = {(g_1^2 + 2\gamma_2^2 + J^2 + 2\Delta_{b_1}^2)}/{3\gamma_2}$, and $g_2=g_1$.}\label{fig6}
 \end{figure}


Figure~\ref{fig6} shows the changes in the real and imaginary parts of the eigenvalues under the asymmetric case for $\gamma_1=2\gamma_2$ and $g_2=g_1$ with Eqs.~(\ref{eq:effective Hamiltonian}) and (\ref{eq:pseudo-Hermitian}). The comparison with Markovian case highlights the addition of green dashed-dotted lines in both the real and imaginary parts, serving to identify the zero eigenvalue and the spectral width feature of the non-Markovian environments, respectively. Interestingly, in the asymmetric case, the non-Markovian-to-Markovian transition shown in Fig.~\ref{fig6}(a) and (c) exhibits not only an EP4-to-EP3 transition but also an EP4-to-EP2 transition, in contrast to the symmetric case Fig.~\ref{fig:Fig3}(c) and (d). Additionally,  Fig.~\ref{fig6}(b) and (d) exhibit an additional green dashed line compared with the Markovian counterpart, corresponding to the chosen spectral width of the non-Markovian environments. In contrast, Fig.~\ref{fig6}(b) and (d) present the asymmetric case: $2{\gamma _2} = {\gamma _1} \ne {\gamma _2}$. In this case, the critical coupling strength at the EP4 is $g_{\mathrm{1EP4}}/2\pi=1.610$ MHz. Compared with Fig.~\ref{fig:Fig3}(a), the difference is that the symmetry of the real part of eigenvalues with respect to the axis ${\rm{Re}}\left[ {\rm E} \right] = {\Delta _a}$ is broken. Moreover, for $g_1<g_{\mathrm{1EP4}}$, the imaginary part, as shown in Fig.~\ref{fig6}(d) compared to Fig.~\ref{fig:Fig3}(b), displays a splitting of the two complex-conjugate pairs.

\begin{figure}[h]
   \centerline{
   \includegraphics[width=8.4cm, height=4.0cm, clip]{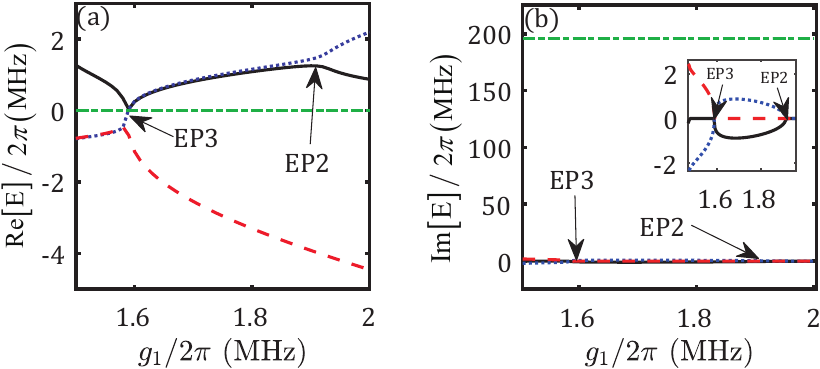}}
   \caption{The evolution of the eigenvalues for the effective Hamiltonian in Eq.~(\ref{eq:effective Hamiltonian}) with Eq. (\ref{eq:pseudo-Hermitian}) as the coupling strength $g_1$ varies at environmental spectral width $\lambda_1/2\pi = \lambda_2 /2\pi= 200$ MHz (Markovian approximation). Herein, (a) and (b) present the real and imaginary parts of these eigenvalues. We set the asymmetry parameters to $\gamma_1/2\pi=2\gamma_2/2\pi=3$ MHz, $\Delta_{b_{2}} = -2\Delta_{b_{1}} = -2 \sqrt{1.04 g_1^2 - \gamma_2^2}$, $J=0.959g_1$, $\Delta_a = 0$, and $g_2=2 g_1$. The remaining parameters are set to $\Gamma_1/2\pi = \Gamma_2 /2\pi= 4.5$ MHz.}\label{fig7}
 \end{figure}

Figure~\ref{fig7}(a) and (b) show the changes in the real and imaginary parts of the eigenvalues during the non-Markovian to Markovian regimes under the asymmetric case with $\gamma_1=2\gamma_2$ and $g_2=2g_1$. The comparison with Markovian case highlights the addition of green dashed-dotted lines in both the real and imaginary parts, serving to identify the zero eigenvalue and the spectral width feature of the non-Markovian environments, respectively. Notably, in the asymmetric-parameter regime, an EP2 emerges that is absent in the symmetric case. This arises from eigenvalue splitting induced by the breaking of parameter-space symmetry.

\section{Details of the classification of higher-order EPs}\label{Classification-details}
Regarding the discussion and analysis of EP5, Appendix~\ref{General case-double} has already been discussed and will not be elaborated further here. In the following discussion, we consider that the environments of the two ports of the cavity $a$ are in non-Markovian regimes, and also the environments of the cavities $b_1$ and $b_2$, where the environments are composed of a series of bosonic modes. The cavity is coupled to the $k$th mode of the non-Markovian environments via the annihilation operators ${e}_k$, ${f}_k$, and their respective creation operators ${e}^{\dag}_k$, ${f}^{\dag}_k$, with the mode eigenfrequencies being $\alpha_k$ and $\beta_k$, respectively. The total non-Markovian Hamiltonian in Eq.~(\ref{eq:H_T}) becomes
\begin{align}
{H}_{T}^{\prime} &= \ \Delta_{a}{a}^{\dagger}{a}+\sum_{j=1,2}\left[\Delta_{b_j}{b}_{j}^{\dagger}{b}_{j}+g_{j}({a}^{\dagger}b_{j}+{a}{b}_{j}^{\dagger})\right]
\nonumber\\&+J({b}_{1}^{\dagger}{b}_{2}+{b}_{1}{b}_{2}^{\dagger})+\sum_{k}\Omega_{k}{c}_{k}^{\dagger}{c}_{k}+\sum_{s}\varpi _s{d}_{s}^{\dagger}{d}_{s}
\nonumber\\&+\sum_{k}\alpha_{k}{e}_{k}^{\dagger}{e}_{k}+\sum_{k}\beta_{k}{f}_{k}^{\dagger}{f}_{k}+i\sum_{k}(C_{k}{a}{c}_{k}^{\dagger}-C_{k}^{*}{a}^{\dagger}{c}_{k})
\nonumber\\&+i \sum_{s}(D_{s}{a}{d}_{s}^{\dagger}-D_{s}^{*}{a}^{\dagger}{d}_{s})+i\sum_{k}(A_{k}{b}_1{e}_{k}^{\dagger}-A_{k}^{*}{b}_1^{\dagger}{e}_{k})
\nonumber\\&+i\sum_{k}(B_{k}{b}_2{f}_{k}^{\dagger}-B_{k}^{*}{b}_2^{\dagger}{f}_{k}),\label{HNM'}
\end{align}
where $A_k$ and $B_k$ denote the interaction strengths for the cavity $b_1$ and cavity $b_2$ with their respective non-Markovian environments, which have frequencies $\alpha_k$ and $\beta_k$, respectively. With Eq.~(\ref{HNM'}), the Heisenberg equation reads
\begin{align}
\frac{d}{dt}{a}(t) =&-i\Delta_{a}{a}(t)-ig_1{b_{1}}(t)
-ig_2{b}_{2}(t)-\sum_{k}C_{k}^{*}{c}_{k}(t)\nonumber\\&-\sum_{s}D_{s}^{*}{d}_{s}(t),\nonumber\\
\frac{d}{dt}{b}_{1}(t) =&-i\Delta_{b_1}{b}_{1}(t)-ig_1{a}(t)-iJ{b_{2}}(t)
-\sum_{k}A_{k}^{*}{e}_{k}(t),\nonumber\\
\frac{d}{dt}{b}_{2}(t) =&-i\Delta_{b_2}{b}_{2}(t)-ig_2{a}(t)-iJ{b_{1}}(t)
-\sum_{k}B_{k}^{*}{f}_{k}(t),\nonumber\\
\frac{d}{dt}{c}_{k}(t) =&-i\Omega_{k}{c}_{k}(t)+C_{k}{a}(t),\label{abcdNonMarkovian'} \\
\frac{d}{dt}{d}_{s}(t) =&-i\varpi _s{d}_{s}(t)+D_{s}{a}(t),\nonumber\\
\frac{d}{dt}{e}_{k}(t) =&-i\alpha_{k}{e}_{k}(t)+A_{k}{b}_1(t),\nonumber\\
\frac{d}{dt}{f}_{k}(t) =&-i\beta_{k}{f}_{k}(t)+B_{k}{b}_2(t). \nonumber
\end{align}
By solving Eq.~(\ref{abcdNonMarkovian'}), we obtain the environmental operators as follows
\begin{equation}
\begin{aligned}
{c}_{k}(t)&={c}_{k}(0)e^{-i\Omega_{k}t}+C_{k}\int_{0}^{t}{a}(\tau)e^{-i\Omega_{k}(t-\tau)}d\tau,
\\{d}_{s}(t)&={d}_{s}(0)e^{-i\varpi _s t}+D_{s}\int_{0}^{t}{a}(\tau)e^{-i\varpi _s (t-\tau)}d\tau,
\\{e}_{k}(t)&={e}_{k}(0)e^{-i\alpha_{k}t}+A_{k}\int_{0}^{t}{b}_1(\tau)e^{-i\alpha_{k}(t-\tau)}d\tau,
\\{f}_{k}(t)&={f}_{k}(0)e^{-i\beta_{k}t}+B_{k}\int_{0}^{t}{b}_2(\tau)e^{-i\beta_{k}(t-\tau)}d\tau,
\label{cdt0NonMarkovian'}
\end{aligned}
\end{equation}
where the first term reflects the free evolution of the non-Markovian environmental fields, while the second term captures the non-Markovian feedback effects from the environments onto the cavities. By substituting Eq.~(\ref{cdt0NonMarkovian'}) into Eq.~(\ref{abcdNonMarkovian'}), we derive the non-Markovian Heisenberg-Langevin equation for the cavity operators
\begin{align}
\frac{d}{dt}{a}(t) =&-i\Delta_{a}{a}(t)-ig_1{b}_{1}(t)-ig_2{b}_{2}(t)+{K}_1(t)+{K}_2(t) \nonumber \\
&-\int_{0}^{t}{a}(\tau)f_1(t-\tau)d\tau-\int_{0}^{t}{a}(\tau)f_2(t-\tau)d\tau, \nonumber\\
\frac{d}{dt}{b}_{1}(t) =&-i\Delta_{b_1}{b}_{1}(t)-ig_1{a}(t)-iJ{b_{2}}(t)+\tilde{K}_1(t) \nonumber \\
&-\int_{0}^{t}{b}_1(\tau)\tilde f_1(t-\tau)d\tau, \label{dotab12NonMarkovian7}\\
\frac{d}{dt}{b}_{2}(t) =&-i\Delta_{b_2}{b}_{2}(t)-ig_2{a}(t)-iJ{b_{1}}(t)+\tilde{K}_2(t) \nonumber\\
&-\int_{0}^{t}{b}_2(\tau) \tilde f_2(t-\tau)d\tau,\nonumber
\end{align}
where 
\begin{align}
&\tilde {K}_1(t)=-\sum_{k}A_{k}^{*}{e}_{k}(0)e^{-i\alpha_{k}t}=\int_{-\infty}^{\infty}\tilde \kappa^{*}_1(t-\tau){b}^{(\rm in)}_1(\tau)d\tau,\nonumber\\
&\tilde {K}_2(t)=-\sum_{k}B_{k}^{*}{f}_{k}(0)e^{-i\beta_{k}t}=\int_{-\infty}^{\infty}\tilde \kappa^{*}_2(t-\tau){b}^{(\rm in)}_2(\tau)d\tau,\nonumber\\
&{b}^{(\rm in)}_1(t)=-\frac{1}{\sqrt{2\pi}}\sum_{k}{e}_{k}(0)e^{-i\alpha_{k}t},\nonumber\\ 
&{b}^{(\rm in)}_2(t)=-\frac{1}{\sqrt{2\pi}}\sum_{k}{f}_{k}(0)e^{-i\beta_{k}t}, \nonumber\\
&\tilde \kappa_{1}(t-\tau)=\frac{1}{\sqrt{2\pi}}\int e^{i\omega(t-\tau)}A(\omega)d\omega,\label{K_k_number}\\
&\tilde \kappa_{2}(t-\tau)=\frac{1}{\sqrt{2\pi}}\int e^{i\omega(t-\tau)}B(\omega)d\omega,\nonumber\\
&\tilde f_{1}(t)=\int \tilde S_{1}(\omega)e^{-i\omega t}d\omega, \tilde f_{2}(t)=\int \tilde S_{2}(\omega)e^{-i\omega t}d\omega,\nonumber\\
&\tilde S_1(\omega)=\sum_{k}|A_{k}|^{2}\delta(\omega-\alpha_{k}), \nonumber
\end{align}
and $\tilde S_2(\omega)=\sum_{k}|B_{k}|^{2}\delta(\omega-\beta_{k})$. Defining $\chi_1{Z}_1(t) = -\tilde{K}_1(t)+\int_{0}^{t}{b_1}(\tau)\tilde f_1(t-\tau)d\tau$, $\chi_2{Z}_2(t) = -\tilde{K}_2(t)+\int_{0}^{t}{b_2}(\tau)\tilde f_2(t-\tau)d\tau$ with $\chi_{1(2)}\equiv\sqrt{{\Lambda _{1(2)} \Upsilon_{1(2)}}/{2}}$, using the non-Markovian input-output relations  ${b}_{1}^{(\rm out)}(t)+{b}_{1}^{(\rm in)}(t)=\int_{0}^{t}\tilde \kappa_{1}(\tau-t){b}_1(\tau)d\tau$, ${b}_{2}^{(\rm out)}(t)+{b}_{2}^{(\rm in)}(t)=\int_{0}^{t}\tilde \kappa_{2}(\tau-t){b}_2(\tau)d\tau
$, and imposing CPA, Eq.~(\ref{dotab12NonMarkovian7}) becomes
\begin{align}
\frac{d}{dt}{a}(t) =&-i\Delta_{c}{a}(t)-ig_1{b_{1}}(t)-ig_2{b}_{2}(t)\nonumber\\
&-\mu_1{Y}_1(t)-\mu_2{Y}_2(t),
\nonumber\\
\frac{d}{dt}{b}_{1}(t) =&-i(\Delta_{1}-i\Upsilon_{1}){b}_{1}(t)-ig_1{a}(t)\nonumber\\
&-iJ{b_{2}}(t) - \chi_1Z_1(t),
\nonumber\\
\frac{d}{dt}{b}_{2}(t) =&-i(\Delta_{2}-i\Upsilon_{2}){b}_{2}(t)-ig_2{a}(t)\nonumber\\
&-iJ{b_{1}}(t) - \chi_2Z_2(t),
\label{dotab12YNonMarkovianE7}\\
\frac{d}{dt}{Y}_1(t)=&\lambda_1{Y}_1(t)+\mu_1{a}(t),
\nonumber\\
\frac{d}{dt}{Y}_2(t)=&\lambda_2{Y}_2(t)+\mu_2{a}(t),
\nonumber\\
\frac{d}{dt}{Z}_1(t)=&\Lambda_1{Z}_1(t)+\chi_1{b_1}(t) ,\nonumber\\
\frac{d}{dt}{Z}_2(t)=&\Lambda_2{Z}_2(t)+\chi_2{b_2}(t).\nonumber
\end{align}
Equation~(\ref{dotab12YNonMarkovianE7}) can be written as a matrix form with $\mathbf{N_7}=({a}, {b}_{1}, {b}_{2}, {Y}_{1}, {Y}_{2}, {Z}_{1}, {Z}_{2})^{T}$
\begin{equation}
\begin{aligned}
\dot{\mathbf{N}}_7=-i{H}_{7}\mathbf{N_7}.
\label{V}
\end{aligned}
\end{equation}
In Eq.~(\ref{V}), the effective Hamiltonian ${H}_{7}$ is written as
\begin{equation}
\begin{aligned}\!\!\!\!\!\!\!
\left( {\begin{array}{*{20}{c}}
{{\Delta _a}}&{{g_1}}&{{g_2}}&{ - i{\mu _1}}&{ - i{\mu _2}}&0&0\\
{{g_1}}&{{\Delta _{{b_1}}} - i{\Upsilon _1}}&J&0&0&{-i{\chi _1}}&0\\
{{g_2}}&J&{{\Delta _{{b_2}}} - i{\Upsilon _2}}&0&0&0&{-i{\chi _2}}\\
{i{\mu _1}}&0&0&{i{\lambda _1}}&0&0&0\\
{i{\mu _2}}&0&0&0&{i{\lambda _2}}&0&0\\
0&{i{\chi _1}}&0&0&0&{i{\Lambda _1}}&0\\
0&0&{i{\chi _2}}&0&0&0&{i{\Lambda _2}}
\end{array}} \right)\!\!\!.\!\!\!
\label{7*7NM}
\end{aligned}
\end{equation}

Building on the Hamiltonian in Eq.~(\ref{7*7NM}), EP7 in the non-Markovian dynamics can be analyzed in direct analogy with the preceding discussions. Moreover, other special cases, such as when cavity $a$ is coupled to distinct non-Markovian reservoirs while cavities $b_1$ and $b_2$ are coupled to the same non-Markovian reservoir, admit a similar analysis. The corresponding effective Hamiltonians $H_{6}$ leading to EP6 can be obtained through an appropriate
reduction of the Hamiltonian in Eq.~(\ref{7*7NM}). For brevity, we do not present the detailed analysis here.

\end{document}